\documentclass[sigconf, nonacm]{acmart}

\newcommand\vldbdoi{XX.XX/XXX.XX}
\newcommand\vldbpages{XXX-XXX}
\newcommand\vldbvolume{14}
\newcommand\vldbissue{1}
\newcommand\vldbyear{2020}
\newcommand\vldbauthors{\authors}
\newcommand\vldbtitle{\shorttitle} 
\newcommand\vldbavailabilityurl{URL_TO_YOUR_ARTIFACTS}
\newcommand\vldbpagestyle{plain} 

\usepackage{color}
\usepackage{ising_v1}
\def\submission{1}
\newcommand{\sys}{PackVFL\xspace}
\usepackage{multirow}
\usepackage{makecell}
\usepackage{bm}
\usepackage{subcaption}
\usepackage[ruled,linesnumbered]{algorithm2e}
\usepackage{stmaryrd}
\usepackage{amsmath}
\usepackage{tikz}
\usepackage{balance}

\begin{document}
\title{\sys: Efficient HE Packing for Vertical Federated Learning}

\author{Liu Yang}
\affiliation{
  \institution{Hong Kong University of Science and Technology}
  \city{Hong Kong}
  \country{China}
}
\email{lyangau@cse.ust.hk}

\author{Shuowei Cai}
\affiliation{
  \institution{Hong Kong University of Science and Technology (Guangzhou)}
  \city{Guangzhou}
  \country{China}
}
\email{scaiak@connect.hkust-gz.edu.cn}

\author{Di Chai}
\affiliation{
  \institution{Hong Kong University of Science and Technology}
  \city{Hong Kong}
  \country{China}
}
\email{dchai@cse.ust.hk}

\author{Junxue Zhang}
\author{Han Tian}
\author{Yilun Jin}
\affiliation{
  \institution{Hong Kong University of Science and Technology}
  \city{Hong Kong}
  \country{China}
}
\email{jzhangcs.htianab.yilun.jin@connect.ust.hk}

\author{Kun Guo}
\affiliation{
  \institution{Fuzhou University}
  \city{Fuzhou}
  \country{China}
}
\email{gukn@fzu.edu.cn}

\author{Kai Chen}
\author{Qiang Yang}
\affiliation{
  \institution{Hong Kong University of Science and Technology}
  \city{Hong Kong}
  \country{China}
}
\email{kaichen.qyang@cse.ust.hk}

\begin{abstract}
  As an essential tool of secure distributed machine learning, vertical federated learning (VFL) based on homomorphic encryption (HE) suffers from severe efficiency problems due to data inflation and time-consuming operations. To this core, we propose~\sys, an efficient VFL framework based on packed HE (PackedHE), to accelerate the existing HE-based VFL algorithms.~\sys packs multiple cleartexts into one ciphertext and supports single-instruction-multiple-data (SIMD)-style parallelism. We focus on designing a high-performant matrix multiplication (MatMult) method since it takes up most of the ciphertext computation time in HE-based VFL. Besides, devising the MatMult method is also challenging for PackedHE because a slight difference in the packing way could predominantly affect its computation and communication costs. Without domain-specific design, directly applying SOTA MatMult methods is hard to achieve optimal.

  Therefore, we make a three-fold design: 1) we systematically explore the current design space of MatMult and quantify the complexity of existing approaches to provide guidance; 2) we propose a hybrid MatMult method according to the unique characteristics of VFL; 3) we adaptively apply our hybrid method in representative VFL algorithms, leveraging distinctive algorithmic properties to further improve efficiency. As the batch size, feature dimension and model size of VFL scale up to large sizes,~\sys consistently delivers enhanced performance. Empirically,~\sys propels existing VFL algorithms to new heights, achieving up to a 51.52$\times$ end-to-end speedup. This represents a substantial 34.51$\times$ greater speedup compared to the direct application of SOTA MatMult methods.
  

\end{abstract}

\maketitle

\pagestyle{\vldbpagestyle}
\begingroup\small\noindent\raggedright\textbf{PVLDB Reference Format:}\\
\vldbauthors. \vldbtitle. PVLDB, \vldbvolume(\vldbissue): \vldbpages, \vldbyear.\\
\href{https://doi.org/\vldbdoi}{doi:\vldbdoi}
\endgroup
\begingroup
\renewcommand\thefootnote{}\footnote{\noindent
This work is licensed under the Creative Commons BY-NC-ND 4.0 International License. Visit \url{https://creativecommons.org/licenses/by-nc-nd/4.0/} to view a copy of this license. For any use beyond those covered by this license, obtain permission by emailing \href{mailto:info@vldb.org}{info@vldb.org}. Copyright is held by the owner/author(s). Publication rights licensed to the VLDB Endowment. \\
\raggedright Proceedings of the VLDB Endowment, Vol. \vldbvolume, No. \vldbissue\ %
ISSN 2150-8097. \\
\href{https://doi.org/\vldbdoi}{doi:\vldbdoi} \\
}\addtocounter{footnote}{-1}\endgroup

\ifdefempty{\vldbavailabilityurl}{}{
\vspace{.3cm}
\begingroup\small\noindent\raggedright\textbf{PVLDB Artifact Availability:}\\
The source code, data, and/or other artifacts have been made available at \url{\vldbavailabilityurl}.
\endgroup
}

\section{Introduction}


Vertical federated learning (VFL)~\cite{yang2019federated,yang2023survey} based on homomorphic encryption (HE)~\cite{acar2018survey} gradually becomes a trend of secure distributed machine learning among data silos~\cite{mcmahan2021advances,chen2021homomorphic,pfitzner2021federated,fink2021artificial}. VFL solves the problem: multiple data owners,~\eg, companies or institutions, hold vertically partitioned data. They want to collaboratively train models without leaking the original data. During the process of HE-based VFL, raw data are maintained locally. Only intermediate results for calculating model updates are encrypted by HE~\cite{zhu2019deep} and exchanged among parties for further cryptographic computation. HE-based VFL is crucial for legally enriching ML features in both academia and industry.


However, state-of-the-art (SOTA) HE-based VFL algorithms~\cite{yang2019federated,chen2021homomorphic,zhang2020additively,cheng2021secureboost} suffer from severe efficiency issues, which hinder VFL's wider application. On the one hand, their adopted HE methods,~\eg, Paillier~\cite{paillier1999public}, largely inflate data size to more than 40$\times$~\cite{zhang2020batchcrypt}, leading to communication and memory overheads. On the other hand, the adopted HE methods introduce time-consuming cryptographic operations, dominating the training process. We dig into these operations and find that matrix multiplication (MatMult) is the efficiency bottleneck. Taking the VFL-LinR algorithm~\cite{yang2019federated} as an example, MatMult gradually occupies the majority of cryptographic computation time, up to 99.23\%, shown in Tab.~\ref{tab:motivation}.

%
%
%
%


\parab{Our Contribution.} In this paper, we propose~\sys, an efficient VFL framework to accelerate the existing HE-based VFL algorithms. In detail,~\sys is based on packed homomorphic encryption (PackedHE)~\cite{brakerski2013packed,brakerski2014leveled,fan2012somewhat,cheon2017homomorphic} and packs multiple cleartexts into one ciphertext to alleviate the data inflation problem and support parallel computation,~\ie, single-instruction-multiple-data (SIMD), over the packed values. More specifically, we focus on devising a high-performant MatMult method tailored for the VFL scenario. Empirically, we show the superiority of~\sys, with 33.30$\times$, 3.22$\times$, 51.52$\times$ speedup over SOTA HE-based VFL algorithms,~\ie, VFL-LinR~\cite{yang2019federated}, CAESAR~\cite{chen2021homomorphic}, VFL-NN~\cite{zhang2020additively} algorithms, respectively, in the end-to-end experiment.

\subsection{Our Techniques}
As the most time-consuming cryptographic operation in HE-based VFL, MatMult is also the design challenge of~\sys. When PackedHE MatMult is adopted in the VFL scenario, the way of packing cleartexts primarily affects its computation and communication cost. Without domain-specific design, MatMult still seriously slows down the training process. For instance, we utilize the naive PackedHE MatMult method~\cite{halevi2014algorithms} in VFL-LinR and find it even slower than Paillier with small batch size,~\eg, 32, as shown in Tab.~\ref{tab:motivation}. The reason is the naive (row-order) method contains too many rotation operations~\cite{halevi2014algorithms}, causing a significant computation overhead. We provide a detailed explanation in~\S\ref{sec:analysis}.

Besides, SOTA PackedHE MatMult methods proposed by~\cite{juvekar2018gazelle,mishra2020delphi,zhang2021gala,huang2022cheetah,hao2022iron} are also sub-optimal for the VFL scenario without domain specific designs. Among them, the methods of GAZELLE~\cite{juvekar2018gazelle}, DELPHI~\cite{mishra2020delphi} and GALA~\cite{zhang2021gala} cause extra computation overheads, while the methods of Cheetah~\cite{huang2022cheetah} and Iron~\cite{hao2022iron} result in extensive communication overheads in VFL. Based on this fact, we raise the question:~\textit{Can we design a PackedHE MatMult method that is ideally tailored for the current HE-based VFL algorithms?} As we will show, the answer is yes. Our new MatMult method outperforms SOTA PackedHE MatMult methods theoretically in~\S\ref{sec:method} and empirically in~\S\ref{sec:experiment}.

\begin{table}
  \centering
\footnotesize    
  \begin{tabular}{c|c|c}
      \toprule
      Batch size & VFL-LinR~\cite{yang2019federated} & Naively applying PackedHE \\
      \midrule
      2 & \bf{\underline{0.014s (10.14\%)}} & 0.037s (75.47\%) \\
      8 & \bf{\underline{0.059s (29.29\%)}} & 0.227s (95.37\%) \\
      32 & \bf{\underline{0.459s (65.19\%)}} & 0.834s (98.73\%) \\
      128 & 5.487s (87.66\%) & \bf{\underline{4.079s (99.76\%)}} \\
      512 & 82.39s (96.71\%) & \bf{\underline{20.73s (99.94\%)}} \\
      2048 & 1419s (99.23\%) & \bf{\underline{111.1s (99.98\%)}} \\
      \bottomrule
  \end{tabular}
  \caption{Time consumed by cryptographic operations in one training batch. Numbers in parentheses represent the proportion of MatMult.}
  \vspace{-0.7cm}

  \label{tab:motivation}
\end{table}



\subsubsection{Systematical Exploration of Design Space.}
As one of the design emphases of PackedHE, the MatMult method has been continuously undergoing updates and iterations~\cite{juvekar2018gazelle,mishra2020delphi,zhang2021gala,huang2022cheetah,hao2022iron,halevi2014algorithms}. Therefore, we provide a comprehensive analysis of the current design space and quantify the complexity of existing approaches to guide our design of~\sys in~\S\ref{sec:analysis}. Based on the main idea of packing cleartexts, we divide them into slot packing methods~\cite{juvekar2018gazelle,mishra2020delphi,zhang2021gala,sav2021poseidon,xu2022hercules,halevi2014algorithms} and coefficient packing methods~\cite{huang2022cheetah,hao2022iron}. Taking the MatMult between cleartext matrix $\bm{X}$ and ciphertext vector $\llbracket \bm{y} \rrbracket$ as as example, we compare their methodologies. Slot packing methods follow the standard encoding procedure of PackedHE. It arranges which cleartexts of $\bm{X}$ are encoded together into a plaintext and their order. On the contrary, coefficient packing methods deviate from the standard procedure and directly map the cleartexts of $\bm{X}$ to specific coefficients of plaintext. The slot and coefficient packing methods have pros and cons considering computation and communication complexities.~\textit{Thus, the challenge lies in selecting the appropriate design path and further driving domain-specific innovations.}

\subsubsection{MatMult Design for VFL Characteristics.}
To overcome the above challenge, we summarize three characteristics of the VFL's required MatMult operation and design a hybrid MatMult method correspondingly in~\S\ref{sec:method}. The first characteristic is that the two operands of VFL MatMult are held by geo-distributed parties. One operand is encrypted by one party and transmitted to another for MatMult computation. After that, the resulting ciphertext is sent back. Through theoretical and empirical analyses of computation and communication costs, we opt for the slot packing concept. As a result, we propose~\textit{\sys's diagonal method} as the core component of our hybrid approach, detailed in~\S\ref{subsec:diagonal}. The second characteristic is wide-range operand size. In VFL, the operand size of MatMult is related to batch size, feature dimension, and model architecture, with a wide range from small to large. Hence, we design delicate~\textit{input packing} and~\textit{input partitioning} techniques to further improve efficiency, correspondingly for the small- and large-operand situation in~\S\ref{subsec:input_pp}. The third characteristic is that the MatMult result is passively decrypted after transmission to the party with secret key. No extra operation is conducted. Due to this characteristic, we design the~\textit{lazy rotate-and-sum (RaS)} mechanism as the last component of our hybrid method to eliminate the remaining time-consuming ciphertext operation (rotation~\cite{halevi2014algorithms}) contained at the end of MatMult in~\S\ref{subsec:lazy_ras}.

Till now, our hybrid method has already shown its superiority over SOTA PackedHE MatMult methods, discussed in~\S\ref{sec:method}. For the computation comparison in~\S\ref{subsec:exp_matmult}, our hybrid method performs the best with the highest 846$\times$ and 1.24$\times$ speedup over the naive method~\cite{halevi2014algorithms} and the SOTA method of GALA~\cite{zhang2021gala}, respectively.


\subsubsection{Adaption Design to SOTA VFL Algorithms.}
To further improve the end-to-end training efficiency, we dig into the secure protocols of three representative HE-based VFL algorithms,~\ie, VFL-LinR~\cite{yang2019federated}, CAESAR~\cite{chen2021homomorphic}, VFL-NN~\cite{zhang2020additively}~\footnote{Our focus on VFL-LinR, CAESAR, and VFL-NN is due to their foundational status and widespread recognition in HE-based VFL applications~\cite{yang2023survey}.} for illustration, and adaptively apply our MatMult method in them with three extra delicate mechanisms, leveraging their distinctive algorithmic properties, in~\S\ref{sec:adaption}. More detailedly, we design the~\textit{multiplication level reduction} mechanism, which modifies CAESAR's original computation process to use more efficient PackedHE parameters in~\S\ref{subsec:caesar}. To make the lazy RaS component of our hybrid method feasible with this modification, we also design the~\textit{cleartext inverse RaS} mechanism. In~\S\ref{subsec:vfl_nn}, focusing on VFL-NN, we encounter a new MatMult scenario between matrices $\bm{X} \llbracket \bm{Y} \rrbracket$. Rather than adhering to the conventional approach to diagonally encode the cleartext matrix $\bm{X}$, we change our mind to diagonally encode the ciphertext matrix $\llbracket \bm{Y} \rrbracket$, significantly enhancing efficiency. Moreover, we develop the~\textit{transposed matrices' diagonal conversion} mechanism, enabling conversion between two encrypted transposed matrices, further halving the communication cost.
 

Together with the adaption optimization,~\sys has more speedup over VFL-LinR~\cite{yang2019federated}, CAESAR~\cite{chen2021homomorphic} and VFL-NN~\cite{zhang2020additively} than the SOTA method of GALA by 9.02$\times$, 0.91$\times$ and 33.1$\times$ respectively. More details are shown in~\S\ref{subsec:exp_e2e}. Besisdes, we also show that none of our designs harms the accuracy of VFL algorithms in~\S\ref{subsec:exp_e2e}.

\subsubsection{\sys's Innovations and New Insights.}
As a cross-disciplinary effort bridging federated learning (FL) and cryptography,~\sys makes significant contributions to both fields. For the FL community,~\sys stands as one of the pioneering works to demonstrate the superiority of PackedHE over Paillier for VFL. We provide a counter-intuitive result that PackedHE is more suitable for VFL with our elaborate design. The VFL's community has the intuition that PackedHE is much slower than Paillier since PackedHE is more complicated and supports more cryptographic operations. We are the first to devise an efficient MatMult method tailored to meet VFL's specific requirements, showcasing its effectiveness across various VFL algorithms.

For the cryptography community, our proposed MatMult method exhibits potential advantages over state-of-the-art slot packing approaches, not only in VFL but also in other related domains such as secure model inference~\cite{mann2022towards}. Additionally, we extend the MatMult scenario from matrix-vector $\bm{X} \llbracket \bm{y} \rrbracket$ to matrix-matrix $\bm{X} \llbracket \bm{Y} \rrbracket$ and design PackedHE techniques of the~\textit{cleartext inverse RaS} and~\textit{transposed matrices' diagonal conversion}, which is unprecedented in earlier works.




 



\subsection{Other Related Work}
\label{sec:related}
Several existing works~\cite{sav2021poseidon,xu2022hercules,ou2020homomorphic,jin2022towards,lu2023squirrel,wang2023xfl} also utilize PackedHE to construct federated algorithms. However,~\cite{ou2020homomorphic,wang2023xfl} provide no design details about MatMult.~\cite{jin2022towards,lu2023squirrel} designed multiplication between ciphertext vectors, which differs from our setting.~\cite{sav2021poseidon,xu2022hercules} involve SOTA PackedHE MatMult designs. But, their target MatMult operation is a single-party continuous MatMult operation without interaction, which is different from the requirements VFL. Therefore, they cannot be adopted in the VFL scenario. Besides, PackedHE hardware acceleration works,~\eg,~\cite{zhang2022sok,samardzic2021f1,su2020fpga}, can also be utilized to improve our performance further since~\sys is a high-level application of PackedHE and changes no basic cryptographic operations.

\section{Background and Preliminaries}
\label{sec:pre}


%


\subsection{Vertical Federated Learning}

Vertical federated learning (VFL)~\cite{yang2019federated, yang2023survey} takes a significant part of cross-silo distributed machine learning, whose participants are business companies~\cite{mcmahan2021advances,fink2021artificial}. Companies usually maintain different portraits for common users,~\ie, vertical data partition~\cite{chen2021homomorphic}. VFL enables secure model training over these abundant distributed features to achieve better prediction accuracy, complying with laws and regulations such as GDPR~\footnote{GDPR is a regulation in EU law on data protection and privacy in the European Union and the European Economic Area. https://gdpr.eu/.}. Therefore, VFL holds substantial real-world importance. For example,~\cite{errounda2022mobility} utilized VFL to help prevent COVID-19~\cite{ilin2021public}.

\begin{figure}[h]
	\centering
	\includegraphics[width=0.7\linewidth]{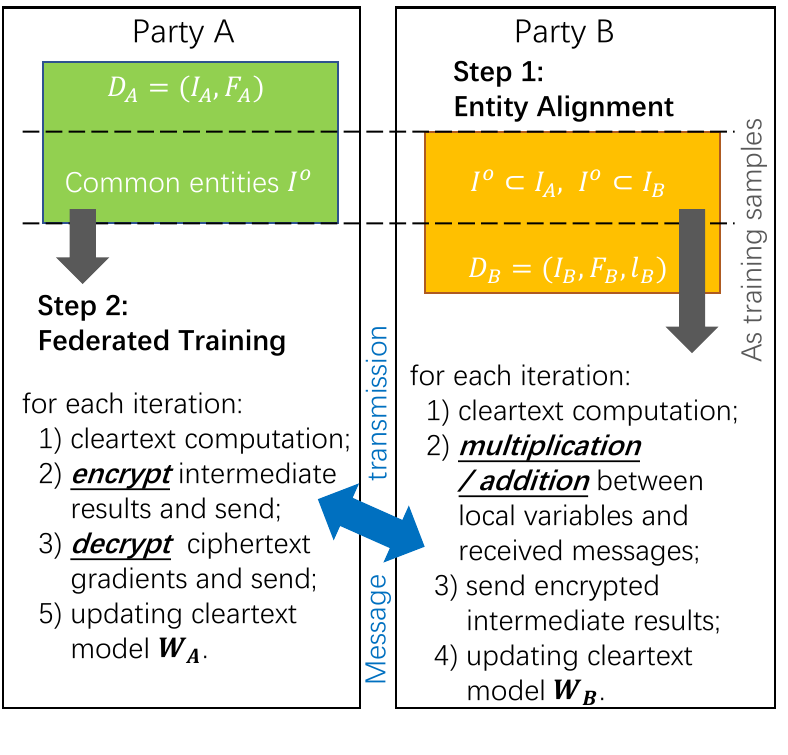}
	\caption{Illustration of HE-based VFL. Some VFL protocols may contain a trusted third party, which we omit to simplify.}
	\label{fig:vfl}
\end{figure}

\subsubsection{HE-based VFL is a Widely-Adopted and Practical Solution.}

During the VFL training process, intermediate results instead of original data are exchanged among parties. To further protect intermediate results from disclosing privacy~\cite{zhu2019deep}, VFL adopts various privacy-preserving methods,~\eg, differential privacy (DP)~\cite{dwork2014algorithmic}, secret sharing (SS)~\cite{beimel2011secret}, and homomorphic encryption (HE)~\cite{acar2018survey}.~\textit{HE-based VFL~\cite{yang2019federated,cheng2021secureboost,chen2021homomorphic,zhang2020additively,cai2022secure} turns out to be a practical solution since DP-based VFL~\cite{xu2019achieving} and SS-based VFL~\cite{mohassel2017secureml,mohassel2018aby3} suffer from either severe accuracy loss~\cite{yang2023survey}~\footnote{DP is more adept at defending the adversarial attacks~\cite{lecuyer2019certified} and membership attacks~\cite{bernau2021comparing}, which are orthogonal to the privacy leakage during the training process.} or large communication overhead~\cite{chen2021homomorphic}.} HE could provide the homomorphism between calculation over ciphertext and calculation over cleartext. HE-based VFL encrypts the transmitted messages, conducts computations on the received ciphertext, and decrypts the results to update the model.

Fig.~\ref{fig:vfl} illustrates a two-party HE-based VFL scenario. Party A ($P_A$) owns a private dataset $\mathcal{D}_A=(\mathcal{I}_A,~\bm{F}_A)$, where $\mathcal{I}$ stands for the sample identifiers and $\bm{F}_A$ represents the features. Party B ($P_B$) holds $\mathcal{D}_B=(\mathcal{I}_B,~\bm{F}_B,~\bm{l}_B)$, where $\bm{l}_B$ means labels. The first step of VFL is entity alignment,~\eg, private set intersection (PSI)~\cite{liang2004privacy}, to securely find the samples with common identifiers between two parties and use them as training data. The second step is federated training. At each iteration, $P_A$ and $P_B$ first conduct cleartext calculation over local data, then exchange encrypted ciphertext for multiplication,~\eg, matrix multiplication (MatMult), addition, and decrypt gradients to update model weights $\bm{W}_A, \bm{W}_B$.

\subsubsection{Paillier's Batching is Limited for VFL.}

Paillier~\cite{acar2018survey} is VFL's most commonly adopted HE approach. It inflates data size and introduces time-consuming operations due to limited support for SIMD operations.~\textit{Although Paillier's batching techniques~\cite{zhang2020batchcrypt,jiang2021flashe} tried to fill several cleartexts into one ciphertext, they do not support the MatMult operation.} As shown in Tab.~\ref{tab:motivation}, MatMult dominates the cryptographic operations of the VFL training process with up to 99.23\% proportion. Therefore, Paillier's batching techniques cannot be utilized in VFL. Besides, Paillier is based on the decisional composite residuosity assumption, which is vulnerable to post-quantum attacks~\cite{shor1999polynomial}. Yet, Paillier remains the predominant HE technique in VFL.

\begin{figure*}
    \centering
    \includegraphics[width=0.7\linewidth]{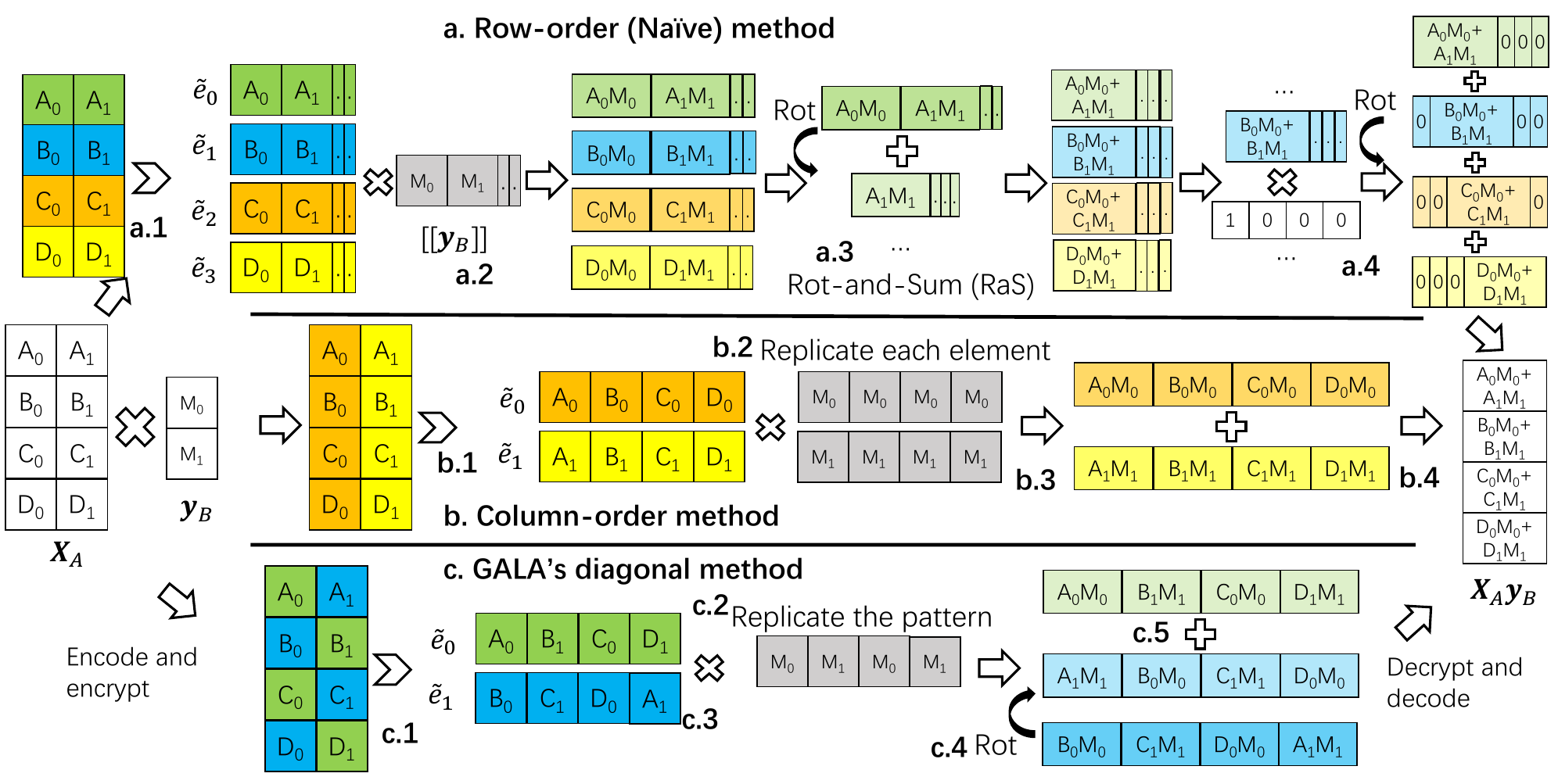}
    \caption{Illustration of slot packing methods,~\ie, the row-order (naive) method, column-order method, and our proposed generalized diagonal method, for PackedHE MatMult operation $\bm{X}_A \llbracket \bm{y}_B \rrbracket = \llbracket \bm{X}_A \bm{y}_B \rrbracket$. We set $m = N' = 4, n = 2$.}
    \label{fig:naive_diagonal}
  \end{figure*}

\subsection{Packed Homomorphic Encryption}
Packed homomorphic encryption (PackedHE),~\eg, BGV~\cite{brakerski2014leveled}, BFV~\cite{fan2012somewhat}, CKKS~\cite{cheon2017homomorphic}, and multiparty BFV~\cite{mouchet2020multiparty}, naturally supports more functional and powerful SIMD-style batching than Paillier. As shown in Fig.~\ref{tab:motivation}, PackedHE with the naive MatMult method has already outperformed Paillier as the number of encrypted cleartexts increases. Moreover, PackedHE is based on a variant of learning with errors (LWE),~\ie, ring LWE (RLWE)~\cite{lyubashevsky2013ideal}, which is quantum-resilient~\cite{nejatollahi2019post}.

LWE encrypts a cleartext vector $\bm{x} \in \mathbb{Z}^N_q$ into a ciphertext $\llbracket \bm{x} \rrbracket = (\bm{c}_0, \bm{c}_1) = (\bm{x}-\bm{A}\bm{s}+\bm{e}, \bm{A})$, where $N$ stands for the vector length, $q$ is the cleartext modulus (large prime number), $\bm{A} \in \mathbb{Z}^{N \times N}_q$ is uniformly sampled, $\bm{s} \in \mathbb{Z}^N_q$ represents the secret key, and $\bm{e} \in \mathbb{Z}^N_q$ is small noise to make the problem hard. The decryption can only be conducted with $\bm{s}$: $\bm{c}_0 + \bm{c}_1\bm{s} = \bm{x} - \bm{A}\bm{s}+\bm{e} + \bm{A}\bm{s} = \bm{x} + \bm{e} \approx \bm{x}$. One disadvantage of LWE is that the size of matrix $\bm{A}$ is quadratic with the vector length $N$, which will cause large communication costs to transmit ciphertext.

PackedHE utilizes RLWE to solve this problem. For a cleartext vector $\bm{x} \in \mathbb{C}_{N/2}$, before encryption, PackedHE encodes it to plaintext,~\ie, an integer polynomial $\tilde{x} \in \mathbb{Z}_q[X]/(X^{N}+1)$. For example, the standard encoding of CKKS utilizes cyclotomic polynomial to construct a one-to-one mapping between the cleartext vector and plaintext polynomial~\cite{cheon2017homomorphic}. Then, PackedHE encrypts plaintext $\tilde{x}$ to ciphertext $\llbracket \bm{x} \rrbracket = (\tilde{c}_0, \tilde{c}_1) = (\tilde{x} - \tilde{a}\tilde{s} + \tilde{e}, \tilde{a})$. Values in the cleartext vector can be regarded to be encoded/encrypted in the corresponding position (slot) of plaintext/ciphertext with the same order. Slot number $N' = N/2$ refers to how many cleartexts can be encrypted in one RLWE ciphertext. The size of polynomial $\tilde{a}$ is linear with $N$, which is much more efficient than LWE. 

We conclude the basic RLWE operations that will be used in the following sections:

\begin{icompact}
	\item \textbf{O1 (Add)}, which represents slot-wise addition,~\eg, $\llbracket \bm{x}\rrbracket + \llbracket \bm{y} \rrbracket$ = $\llbracket \bm{x} + \bm{y} \rrbracket$ or $\bm{x} + \llbracket \bm{y} \rrbracket$ = $\llbracket \bm{x} + \bm{y} \rrbracket$;
	\item \textbf{O2 (Mult)}, which represents slot-wise multiplication,~\eg, $\llbracket \bm{x} \rrbracket \times \llbracket \bm{y} \rrbracket = \llbracket \bm{x} \times \bm{y} \rrbracket$ or $\bm{x} \times \llbracket \bm{y} \rrbracket = \llbracket \bm{x} \times \bm{y} \rrbracket$;
	\item \textbf{O3 (Rot)}, which represents a rotation operation that shifts ciphertext slots in sequence. $\mbox{RotL/R}(\llbracket \bm{x} \rrbracket, i)$ denotes rotating $\llbracket \bm{x} \rrbracket$ to the left/right for $i$ position. Rotation is the most costly basic RLWE operation, which is more than 10$\times$ slower than \textbf{O1 (Add)} and \textbf{O2 (Mult)};
	\item \textbf{O4 (HstRot)}, which represents hoisting rotation that is a more efficient optimization to conduct multiple rotation operations over the same ciphertext~\cite{juvekar2018gazelle}. $\mbox{HstRotL/R}(\llbracket \bm{x} \rrbracket, i)$ denotes $i$-position left/right HstRot of $\llbracket \bm{x} \rrbracket$.
\end{icompact}




\subsubsection{Designing PackedHE MatMult is Non-Trivial.}

PackedHE has already achieved good efficiency when conducting basic~\textbf{O1 (Add)} and~\textbf{O2 (Mult)} operations with SIMD property. However, there exist difficulties when designing a more complicated PackedHE MatMult method~\cite{juvekar2018gazelle}.~\textit{The main reason is that PackedHE needs~\textbf{O3 (Rot)} operations to sum values in different slots of ciphertexts. Since matrix multiplication needs a large number of vector inner sum operations, PackedHE MatMult is inclined to be inefficient without proper designs.}

\begin{table*}[h]
  \centering
  \setlength{\tabcolsep}{4pt}
  \footnotesize
  \begin{tabular}{c|c|cccc|cc}
      \toprule
      \multirow{2}{*}{Category} & \multirow{2}{*}{Method} & \multicolumn{4}{c}{\underline{Computation Complexity}} & \multicolumn{2}{c}{\underline{Communication complexity}} \\
       & & \# O1 (Add) & \# O2 (Mult) & \# O3 (Rot) & \# O4 (HstRot) & $P_B$ to $P_A$ & $P_A$ to $P_B$ \\
      \midrule
      \multirow{6}{*}{\makecell[c]{Slot\\packing}} & Naive~\cite{halevi2014algorithms} & $m \log_2 n + m - 1$ & $2m$ & $m \log_2 n + m - 1$ & $0$ & $1$ RLWE-ct & $1$ RLWE-ct \\
       & Column-order~\cite{halevi2014algorithms} & $n - 1$ & $n$ & $0$ & $0$  & $n$ RLWE-ct & $1$ RLWE-ct \\
  & GALA~\cite{zhang2021gala}'s Diagonal & $\min (m, n) - 1 + \log_2 \lceil \frac{n}{m} \rceil$ & $\min (m, n)$ & $\min (m, n) - 1$ & $\log_2 \lceil \frac{n}{m} \rceil $ & $1$ RLWE-ct & $1$ RLWE-ct \\
   & \sys's Diagonal & $\min (m, n) - 1 + \log_2 \lceil \frac{n}{m} \rceil$ & $\min (m, n)$ & $\log_2 \lceil \frac{n}{m} \rceil $ & $\min (m, n) - 1$ & $1$ RLWE-ct & $1$ RLWE-ct \\
   & GALA~\cite{zhang2021gala} & $\lceil \frac{mn}{N'} \rceil - 1$ & $\lceil \frac{mn}{N'} \rceil$ & $\lceil \frac{mn}{N'} \rceil - 1$ & $0$  & $1$ RLWE-ct & $1$ RLWE-ct \\
   &\underline{\textbf{\sys}} & \underline{$\bf{\lceil \frac{mn}{N'} \rceil - 1}$} & \underline{$\bf{\lceil \frac{mn}{N'} \rceil}$} & \underline{$\bf{0}$} & \underline{$\bf{\lceil \frac{mn}{N'} \rceil - 1}$} & \underline{$\bf{1}$ \textbf{RLWE-ct}} & \underline{$\bf{1}$ \textbf{RLWE-ct}} \\
  \midrule
  \makecell[c]{Coefficient\\packing} & Cheetah~\cite{huang2022cheetah} & $0$ & $1$ & $0$ & $0$ & $1$ RLWE-ct & $m$ LWE-ct \\
      \bottomrule
  \end{tabular}
  \caption{Complexity of MatMult methods between matrix and vector. Computation complexity is mainly decided by the numbers (\#) of involved O3 (Rot) and O4 (HstRot) operations. Communication complexity contains messages from $P_B$ to $P_A$ and from $P_A$ to $P_B$. We assume $N$ is large enough with no need for matrix/vector partition operation.}
  \label{tab:complexity}
  \vspace{-0.7cm}
\end{table*}

\section{Systematical Exploration of Design Space}
\label{sec:analysis}

The existing works~\cite{halevi2014algorithms,juvekar2018gazelle,mishra2020delphi,zhang2021gala,huang2022cheetah,hao2022iron} have proposed SOTA designs for PackedHE MatMult. Taking the VFL's MatMult between matrix and vector as an example, we introduce and compare their methods to provide guidance for our design. The VFL's MatMult is formulated as:


\begin{equation}
\label{eq:matmult}
	\bm{X}_A \llbracket \bm{y}_B \rrbracket = \llbracket \bm{X}_A \bm{y}_B \rrbracket,
\end{equation}
where $\bm{X}_A \in \mathbb{R}^{m \times n}$ is held by $P_A$, $\bm{y}_B \in \mathbb{R}^{n \times 1}$ is held by $P_B$. In VFL's process, $P_B$ encrypts $\bm{y}_B$ as $\llbracket \bm{y}_B \rrbracket$ and sends it to $P_A$ for MatMult. The resulting ciphertext is transmitted back to $P_B$ for decryption. $\bm{X}_A$ should also be encoded to plaintexts before MatMult.

We mainly divide the existing MatMult methods of~\cite{halevi2014algorithms,juvekar2018gazelle,mishra2020delphi,zhang2021gala,huang2022cheetah,hao2022iron} into two categories below and will show their advantages and disadvantages in the following of this section:

\begin{icompact}
	\item \textbf{Slot packing methods} that follow the standard encoding procedure of PackedHE, decide which cleartexts are encoded together into a plaintext and arrange their order~\cite{halevi2014algorithms,juvekar2018gazelle,mishra2020delphi,zhang2021gala};
	\item \textbf{Coefficient packing methods} that deviate from the standard procedure and focus on designing novel mapping between cleartexts to specific coefficients of plaintext~\cite{huang2022cheetah,hao2022iron}.
\end{icompact}

In the following of this paper, we mention GAZELLE~\cite{juvekar2018gazelle}, DELPHI~\cite{mishra2020delphi}, GALA~\cite{zhang2021gala}, Cheetah~\cite{huang2022cheetah}, and Iron~\cite{huang2022cheetah} to only represent their MatMult methods for simplicity. Since their primary focus is on secure CNN inference, other technical contributions,~\eg, secure convolution techniques, are not applicable in our VFL scenario. Besides, we assume that $m, n, N$ are both power of two. In the real-world scenario, $m, n$ are not always the power of two, we can conduct the padding operation over $\bm{X}_A$ and $\bm{y}_B$ with zero~\cite{benaissa2021tenseal}. We set $m = 4, n = 2$ for demonstration. Matrix $\bm{X}_A$ has four rows,~\ie, $[A_0, A_1]$, $[B_0, B_1]$, $[C_0, C_1]$, and $[D_0, D_1]$, while column-vector $\bm{y}_B$ contains two elements,~\ie, $M_0$ and $M_1$. $A_i$, $B_i$, and $M_i$ are scalars to illustrate the computation process. Tab.~\ref{tab:complexity} shows both the computation and communication complexity of the involved method. For a clear comparison, we assume $N$ is large enough without needing matrix/vector partition.

\subsection{Slot Packing}
The fundamental idea of slot packing is deciding the encoding logic of the matrix operand~\cite{halevi2014algorithms}. Among the choices, row-order and column-order encodings are relatively intuitive. The slot packing methods are illustrated in Fig.~\ref{fig:naive_diagonal} with $N' = 4$. 



\subsubsection{Row-Order (Naive) Method.} 
The row-order (naive) method~\cite{halevi2014algorithms} is regarded as the most intuitive solution, which is frequently discussed as a baseline in related works~\cite{juvekar2018gazelle,mishra2020delphi,zhang2021gala}. And we also use the naive method to conduct the motivation experiments in Tab.~\ref{tab:motivation}. The efficiency of the naive method is largely slowed down by the required $m \log_2 n + m - 1$~\textbf{O3 (Rot)} operations, shown in Tab.~\ref{tab:complexity}. In our experiments,~\textbf{O3 (Rot)} consumes around 90\% time of the naive method when $m = n = 128$.


Shown in Fig.~\ref{fig:naive_diagonal}, the process of the naive method is: 1) each row of matrix $\bm{X}_A$ is encoded to $\tilde{e}_i$, respectively. For example, $\tilde{e}_0$ contains $[A_0, A_1,~\cdot,~\cdot]$, where we make omissions for vacant slots; 2) we multiply each $\tilde{e}_i$ with ciphertext $\llbracket \bm{y}_B \rrbracket$; 3) since the values in one ciphertext cannot be directly summed up in PackedHE, we conduct $\log_2 n$ Rotation-and-Sum (RaS) for each resulting ciphertext, which shifts the ciphertext for some specific position,~\ie,~\textbf{O3 (Rot)}, and adds the new ciphertext back to the old one. For instance, we conduct $\mbox{RotL}(\llbracket [A_0M_0, A_1M_1, \cdot, \cdot] \rrbracket, 1)$, obtain $\llbracket [A_1M_1, A_0M_0, \cdot, \cdot] \rrbracket$ and conduct $\llbracket [A_0M_0, A_1M_1, \cdot, \cdot] \rrbracket + \llbracket [A_1M_1, A_0M_0, \cdot, \cdot] \rrbracket$. After $m \log_2 n$ RaS containing $m \log_2 n$~\textbf{O3 (Rot)}, we get the result of each row's dot product with $\llbracket \bm{y}_B \rrbracket$ at the first slot; 4) we want to obtain the expected result of MatMult in one ciphertext. To achieve this goal, we multiply each resulting ciphertext of the previous step with a cleartext indicator vector, which is only set to one at the first position and zero at other positions. Finally, we conduct extra $m - 1$~\textbf{O3 (Rot)} to further adjust the maintained values to appropriate slots and sum them together.








\subsubsection{Column-Order Method.} 
The column-order method~\cite{halevi2014algorithms} is less studied, even though owning a small computation complexity with no~\textbf{O3 (Rot)} operations, shown in Tab.~\ref{tab:complexity}. However, it has a nearly $n\times$ communication complexity than naive method. Column-order method follows an opposite logic to the naive method. Fig.~\ref{fig:naive_diagonal} shows its process: 1) each column of matrix $\bm{X}_A$ is encoded into $\tilde{e}_i$. $\tilde{e}_0$ contains $[A_0, B_0, C_0, D_0]$ and $\tilde{e}_1$ contains $[A_1, B_1, C_1, D_1]$; 2) the vector $\bm{y}$ is encoded and encrypted to $n=2$ ciphertexts. Each one separately replicates the corresponding element of $\bm{y}_B$,~\ie, $\llbracket [M_0, M_0, M_0, M_0] \rrbracket$ and $\llbracket [M_1, M_1, M_1, M_1] \rrbracket$. These $n$ RLWE-ciphertexts are sent from $P_B$ to $P_A$, causing large communication overhead. By contrast, the naive method only transmits one RLWE-ciphertext (ct); 3) slot-wise multiplication is conducted for each $\tilde{e}_i$ and corresponding ciphertext from $P_B$; 4) $P_A$ performs slot-wise addition for results of the previous step and sends the obtained one RLWE-ct back to $P_B$.

\subsubsection{GALA's Diagonal Method.} Diagonal methods~\cite{mishra2020delphi,juvekar2018gazelle,zhang2021gala} are popular for slot packing. Their main idea is to place the values required to be added in the same slot of different ciphertexts to reduce the needed~\textbf{O3 (Rot)}. DELPHI~\cite{mishra2020delphi} is a GAZELLE~\cite{juvekar2018gazelle} variant, assuming the input matrix of MatMult can be known in advance and moving part of the computation to the preprocessing phase. Since this assumption is not applicable in VFL, DELPHI degenerates to GAZELLE. GALA~\cite{zhang2021gala} outperforms GAZELLE by making the number of~\textbf{O3 (Rot)} disproportional to $N'$ and eliminating all~\textbf{O4 (HstRot)} on $\llbracket \bm{y}_B \rrbracket$. Therefore, we illustrate the SOTA diagonal method of GALA.

Fig.~\ref{fig:naive_diagonal} shows the process of GALA's diagonal method: 1) each diagonal of matrix $\bm{X}_A$ is encoded into $\tilde{e}_i$, following the diagonal order. $\tilde{e}_0$ contains $[A_0, B_1, C_0, D_1]$ and $\tilde{e}_1$ contains $[B_0, C_1, D_0, A_1]$; 2) the vector $\bm{y}_B$ is encoded and encrypted, replicating the whole pattern, as $\llbracket [M_0, M_1, M_0, M_1] \rrbracket$. This RLWE-ct is sent from $P_B$ to $P_A$; 3) slot-wise multiplication is conducted by $P_A$ for each $\tilde{e}_i$ and the received cipheretext; 4) $P_A$ rotates $\llbracket [B_0 M_0, C1 M_1, D_0 M_0, A_1 M_1] \rrbracket$ to $\llbracket [A_1 M_1, B_0 M_0, C_1 M_1, D_0 M_0] \rrbracket$; 5) $P_A$ conducts slot-wise addition for the result of step (4) and $\llbracket [A_0 M_0, B_1 M_1, C_0 M_0, D_1 M_1] \rrbracket$. The obtained RLWE-ct is sent back to $P_B$ for decryption. Shown in Tab.~\ref{tab:complexity}, GALA's diagonal method maintains a small commnunication complexity same as the naive method as well as achieve a smaller computation complexity.



\subsection{Coefficient Packing}
Coefficient packing methods~\cite{huang2022cheetah,hao2022iron} are recently proposed as SOTA methods, which do not follow previous standard encoding procedures,~\eg,~\cite{brakerski2014leveled,fan2012somewhat,cheon2017homomorphic} and directly map cleartexts to coefficients of a plaintext polynomial. Cheetah~\cite{huang2022cheetah} is proposed as a MatMult method between matrix and vector. Iron~\cite{hao2022iron} extends it to MatMult between matrices. Their computation is only one multiplication, with no other operation,~\eg,~\textbf{O3 (Rot)}. However, they also suffer from high communication costs when applied in the VFL scenario, shown in Tab.~\ref{tab:complexity}. For illustration, we only introduce Cheetah~\cite{huang2022cheetah}, which suits our task,~\ie, Eq.~\ref{eq:matmult}. The process of Cheetah is: 1) $\tilde{X}_A$ is constructed according to mapping $\tilde{X}_A[i \cdot n + n - 1 - j] = \bm{X}_A[i, j]$:

\begin{equation}
\begin{aligned}
	\tilde{X}_A = & A_1 X^0 + A_0 X^1 + B_1 X^2 + B_0 X^3 + \\
	& C_1 X^4 + C_0 X^5 + D_1 X^6 + D_0 X^7 \in \mathbb{Z}_q[X]/(X^{16}+1);
\end{aligned}
\end{equation}
2) $\tilde{y}_B$ is constructed according to mapping $\tilde{y}_B[i] = \bm{y}[i]$:
\begin{equation}
	\tilde{y}_B = M_0 X^0 + M_1 X^1 \in \mathbb{Z}_q[X]/(X^{16}+1);
\end{equation}
3) multiplication is conducted between $\tilde{X}_A$ and $\tilde{y}_B$ to obtain~\footnote{We only show the multiplication between polynomials for better illustration, which is same as~\cite{huang2022cheetah}. In fact, $\tilde{y}_B$ should be encrypted as RLWE-ct before multiplication.}:
\begin{equation}
\begin{aligned}
	\tilde{X}_A \tilde{y}_B = & \dots + (A_0 M_0 + A_1 M_1) X^1 + \dots + (B_0 M_0 + B_1 M_1) X^3 + \\
	& \dots + (C_0 M_0 + C_1 M_1) X^5 + \dots + (D_0 M_0 + D_1 M_1) X^7 + \\
	& \cdots \mod (X^{16}+1, q);
\end{aligned}
\end{equation}
4) the MatMult results can be found with mapping $(\bm{X}_A \bm{y}_B)[i] = (\tilde{X}_A \tilde{y}_B)[i \cdot n + n - 1]$. Cheetah extracts specific RLWE coefficients to separate $m$ LWE-cts using~\cite{chen2021efficient} to protect the privacy of the other unneeded coefficients. In our example, four LWE-cts are extracted from coefficients $A_0 M_0 + A_1 M_1, B_0 M_0 + B_1 M_1, C_0 M_0 + C_1 M_1, D_0 M_0 + D_1 M_1$ and sent to $P_B$ for decryption. As described in~\S\ref{sec:pre}, we can calculate that these extracted LWE-cts are $m(1 + N)/2N\times$ larger than the original RLWE-ct, causing Large communication overheads.


\section{MatMult Design for VFL Characteristics}
\label{sec:method}

\begin{figure}[h]
	\centering
	\includegraphics[width=0.8\linewidth]{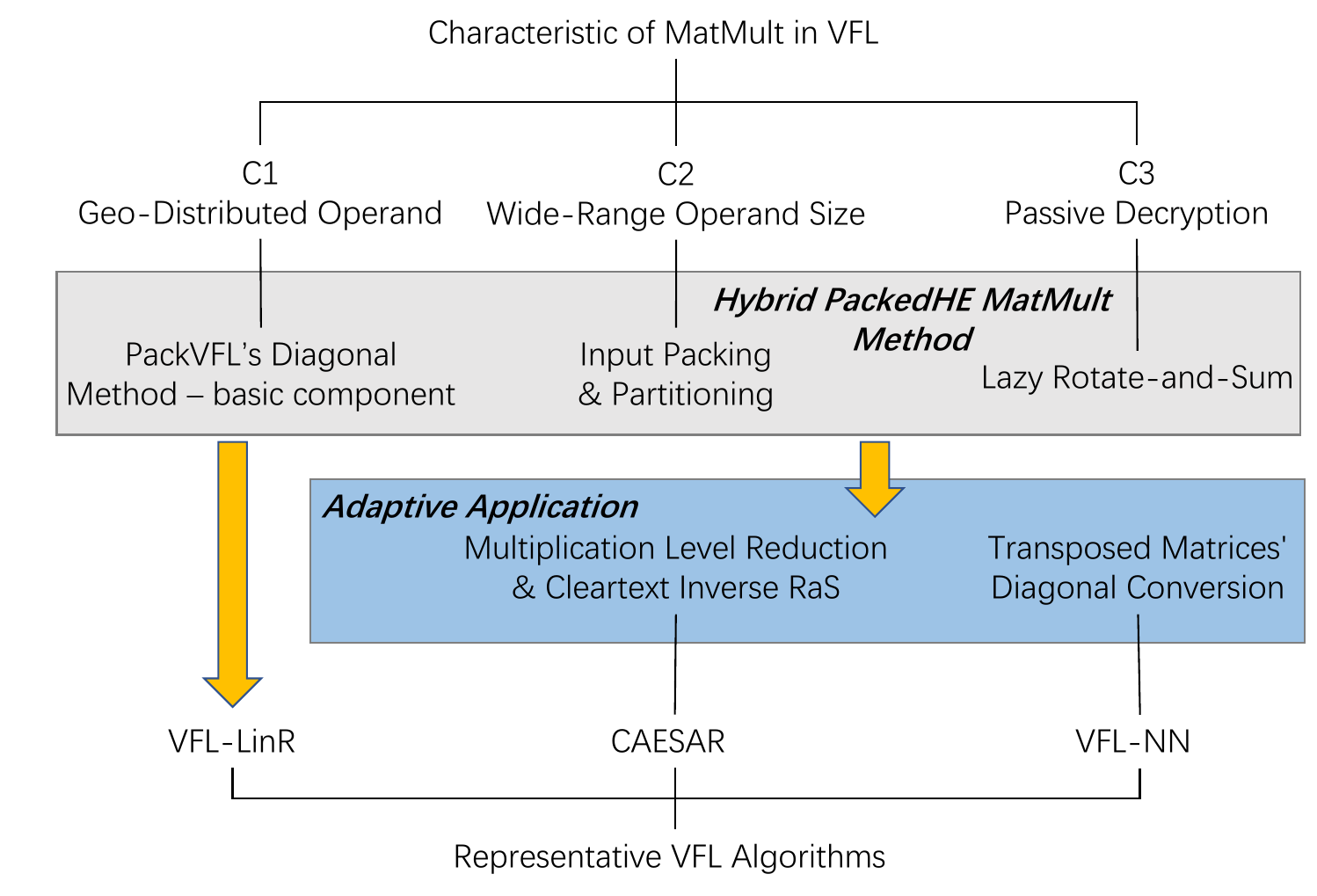}
	\caption{Overview of~\sys.~\sys contains two parts: 1) we design a hybrid MatMult method in terms of the characteristics of VFL; 2) we adaptively apply the proposed MatMult method to representative VFL algorithms.}
	\label{fig:method_overview}
\end{figure}

With the guidance of comprehensive analysis in~\S\ref{sec:analysis}, we conclude characteristics of VFL's MatMult operation and design a high-performant hybrid MatMult method accordingly, shown in Fig.~\ref{fig:method_overview}. The characteristics of VFL's MatMult are summarized below:

	\begin{icompact}
		\item~\textbf{C1 (Geo-Distributed Operand)}: the two operands are owned by geo-distributed parties of VFL. The encrypted operand is sent to the other party for MatMult, and the ciphertext result is transmitted back for decryption;
		\item~\textbf{C2 (Wide-Range Operand Size)}: as a machine learning technique, VFL often has varying batch sizes or feature dimensions, leading to varying operand sizes of MatMult, from large to small;
		\item~\textbf{C3 (Passive Decryption)}: the resulting ciphertexts of MatMult are transmitted to the party owning secret key for pure decryption without extra operations.
	\end{icompact}


\subsection{\sys's Diagonal Method}
\label{subsec:diagonal}
The~\textbf{C1 (Geo-Distributed Operand)} characteristic indicates a wide area network (WAN) situation with relatively small bandwidth. Large communication overheads of MatMult could severely slow down the training process. Besides, parties (companies) of VFL usually contain rich computing resources.~\textit{Our design principle is: trying to reduce the computation complexity after guaranteeing that communication complexity is small enough.} 

\subsubsection{Our Choice of Diagonal Method.}
Considering the~\textbf{C1 (Geo-Distributed Operand)} characteristic of VFL, we choose the diagonal method as our basic component. More specifically, comparing to the naive method, diagonal method can achieve less computation complexity while maintain the same communication complexity. For comparison with column-order and coefficient packing methods, we conduct an experiment where the bandwidth between $P_A$ and $P_B$ is 50MB/s~\cite{jiang2021flashe} and $m = n = 512, N = 8192$. The transmission time of the column-order method and Cheetah is beyond 10s, which is much larger than the computation overhead (under 1s) of diagonal method. Therefore, we choose the most suitable design path for the VFL MatMult.





However, through the systematical analysis in~\S\ref{sec:analysis}, we observe that the SOTA diagonal method of GALA is not optimal for computation efficiency. The reason is that GALA still contains unnecessary~\textbf{O3 (Rot)} operations caused by its diagonal encoding way. Therefore, we propose~\sys's diagonal method, which adopts a different diagonal encoding to further replace the unnecessary~\textbf{O3 (Rot)} with more efficient~\textbf{O4 (HstRot)} operations. 

\begin{figure}[h]
  \centering
  \includegraphics[width=0.7\linewidth]{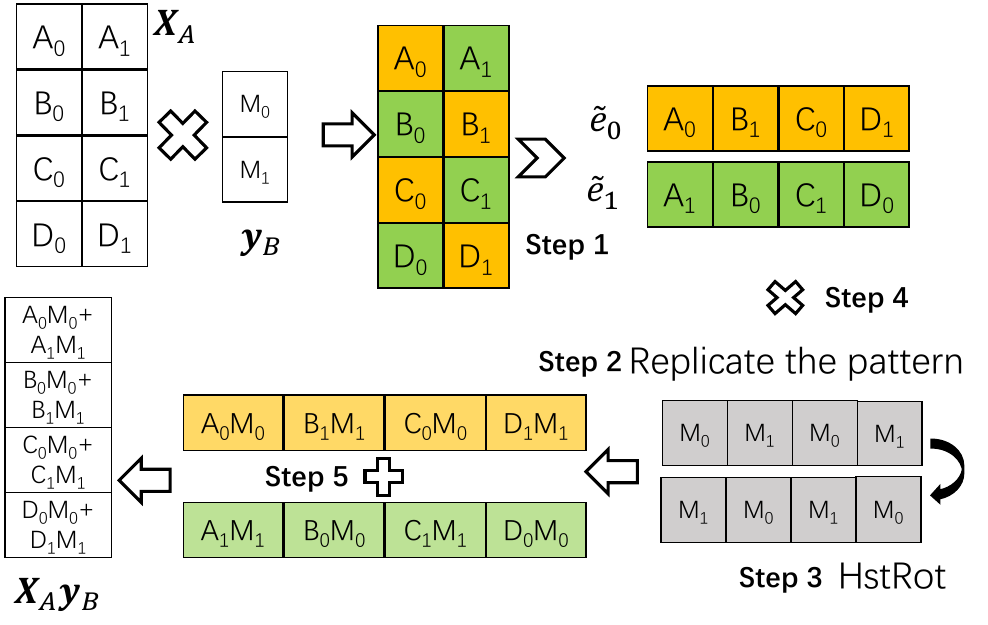}
  \caption{Illustration of \sys's diagonal method.}
  \label{fig:generalized_diagonal}
  \vspace{-0.5cm}
\end{figure}

\sys's diagonal method encodes the matrix $\bm{X}_A$ into:

\begin{equation}
\label{eq:diagonal}
	\tilde{e}_i [j] = \bm{X}_A [j \hspace{-0.2cm} \mod m, \  (i+j) \hspace{-0.2cm} \mod n],
\end{equation}
where $i = 0, 1, ..., \min(m,n)-1$ and $j = 0, 1, ..., \max(m, n)-1$. This equation is a general solution for both $m \leq n$ (short-and-wide matrix) and $m > n$ (tall-and-skinny matrix) scenario. Fig.~\ref{fig:generalized_diagonal} illustrates the tall-and-skinny matrix case of $m > n$. The process is described as: 1) each $\tilde{e}_i$ is generated along the diagonal, following Eq..~\ref{eq:diagonal}. For example, $[A_0, B_1, C_0, D_1]$ are encoded into $\tilde{e}_0$; 2) the ciphertext $\llbracket \bm{y}_B \rrbracket$ is constructed by replicating the whole $y_B$ for $\frac{m}{n}$ times; 3) $n - 1$~\textbf{O3 (Rot)} need to be conducted on $\llbracket \bm{y}_B \rrbracket$ to obtain $n - 1$ new ciphertexts. We use~\textbf{O4 (HstRot)} to replace them for better efficiency; 4) multiplication between $\tilde{e}_i$ and corresponding rotated ciphertext is performed; 5) we add the resulting ciphertexts to get $\llbracket \bm{X}_A \bm{y}_B \rrbracket$. For the short-and-wide matrix case $m \leq n$, the construction of $\llbracket \bm{y}_B \rrbracket$ needs no replication operation and additional RaS operations should be conducted. Readers could refer to Fig.~\ref{fig:hetero_linr_packing} ($m = n$) and Fig.~\ref{fig:caesar_packing} ($m < n$) for more details.








Comparing to GALA in Tab.~\ref{tab:complexity},~\sys's diagonal method substitutes $\min (m, n) - 1 - \log_2 \lceil \frac{n}{m} \rceil$~\textbf{O3 (Rot)} with the same number of~\textbf{O4 (HstRot)} operations generally, achieving the optimal computation complexity. $\lceil \cdot \rceil$ stands for ceiling function. In the case of $m \geq n$,~\sys's diagonal method could substitute all~\textbf{O3 (Rot)} with~\textbf{O4 (HstRot)}, while, in the case of $m < n$, our diagonal method still needs $\log_2 \frac{n}{m}$~\textbf{O3 (Rot)} for the final RaS operations.

\subsection{Input Packing and Partitioning}
\label{subsec:input_pp}
The characteristic~\textbf{C2 (Wide-Range Operand Size)} indicates two scenarios: small size operands ($m, n < N'$) and large size operand ($m > N'$ or $n > N'$). The challenges of efficiently supporting small- and large-size operands are different. The former is to effectively leverage vacant slots for high-performant SIMD, while the latter one is to minimize redundant overheads across blocks as many as possible. We design input packing and partitioning, respectively.


\subsubsection{Input Packing.} 
For scenario $m, n < N'$, we pack multiple matrix diagonals into one plaintext polynomial to fully utilize the slots. In the beginning, we obtain the transformed matrix $\bm{X}' \in \mathbb{R}^{m' \times N'}$ with $N' / n$ sub-matrices vertically concatenated:

\begin{equation}
\label{eq:input_packing}
	\bm{X}' [i \hspace{-0.2cm} \mod m', \ j + n \lfloor \frac{i}{m'} \rfloor ] = \bm{X} [i, j],
\end{equation}
where $m' = \lceil \frac{mn}{N'} \rceil$, $\lfloor \cdot \rfloor$ stands for floor function, $i = 0,1,...,m-1$ and $j = 0,1,...,n-1$. Then, we conduct diagonal encoding for $\bm{X}'$, which is slightly different from Eq.~\ref{eq:diagonal}. We illustrate it in Fig.~\ref{fig:hetero_linr_packing} to compute $\bm{X}_A^T \llbracket \bm{d} \rrbracket$. After we move the lower part of matrix $\bm{X}_A^T$ to the right and form a transformed matrix at step 2, the diagonal encoding is performed in parts. For example, at step 3, $B_0$ is placed after $A_3$, not at the last slot of $\tilde{e}_1$, while $D_0$ is set at the last slot. On the other side, the encrypted vector $\llbracket \bm{d} \rrbracket$ is expanded by replicating its pattern as $\llbracket [M_0, M_1, M_2, M_3, M_0, M_1, M_2, M_3] \rrbracket$, at step 1.










\subsubsection{Input Partitioning.} 
For the scenarios $m > N'$ or $n > N'$, we split the original matrix/vector into multiple blocks in slot size $N'$ and design tricks to further improve efficiency. 

\parab{Large Operand Size is Common in VFL.} VFL's operand sizes are influenced by factors such as batch size, feature dimension, and the number of model parameters. These elements are increasingly scaling up in real-world applications. For instance, within the realm of machine learning, VFL is progressively adopting larger batch sizes — such as 15,000~\cite{keskar2016large}, 4,096~\cite{hoffer2017train}, and 8,192~\cite{lin2020extrapolation} — to enhance parallelism. Besides, the era of big data sees a surge in datasets with high feature dimensions, such as 22,283, 19,993~\cite{liu2017deep}, reflecting a growing trend~\cite{zhai2014emerging}. In addition, as GPT~\cite{brown2020language} gains widespread attention, large models, such as those with 65 billion~\cite{touvron2023llama} and 130 billion parameters~\cite{zeng2022glm}, are becoming a focal point of research.


In the case $m > N' \geq n$, we horizontally split the original matrix $\bm{X}$ into $\frac{m}{N'}$ sub-matrices, then conduct the diagonal method between each sub-matrix and $\llbracket \bm{y}_B \rrbracket$ separately. Besides, we only need to complete one group of~\textbf{O4 (HstRot)} on $\llbracket \bm{y}_B \rrbracket$ for all sub-matrices. For the case $m \leq N' < n$, we provide an example in Fig.~\ref{fig:caesar_packing} to compute $\bm{X}_A \llbracket \langle \bm{w}_A \rangle_2 \rrbracket$. $<\cdot>$ stands for the secret share. At step 1\&2, we vertically divide both the original matrix $\bm{X}_A$ and vector $\langle \bm{w}_A \rangle_2$ into $\frac{n}{N'}$ blocks. At step 3, we conduct diagonal method for each block. Originally, we will obtain $\frac{n}{N'}$ ciphertext results and send them to $P_B$ for decryption, which vastly increases communication complexity. However, we bring forward the addition operation of results in different blocks at step 4. Hence, we need only one transmission of the aggregated ciphertext.





\subsection{Lazy Rotate-and-Sum.}
\label{subsec:lazy_ras}
\begin{figure*}
  \centering
  \begin{subfigure}[b]{0.31\textwidth}
      \centering
      \includegraphics[width=\textwidth]{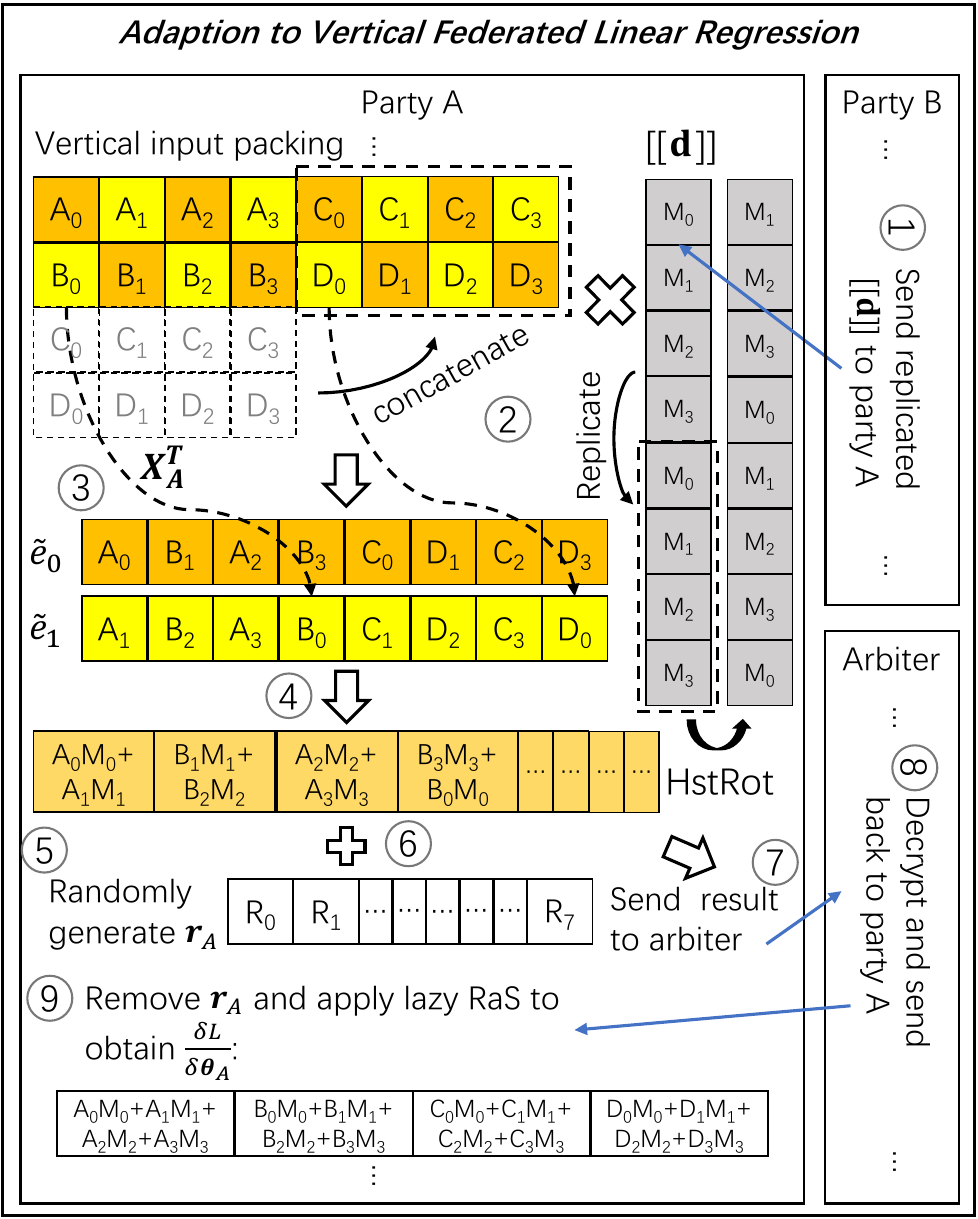}
      \caption{The VFL-LinR algorithm. We set $m = n = 4, N' = 8$. The input packing and lazy RaS techniques are also adopted.}
      \label{fig:hetero_linr_packing}
  \end{subfigure}
  \hfill
  \begin{subfigure}[b]{0.31\textwidth}
      \centering
      \includegraphics[width=\textwidth]{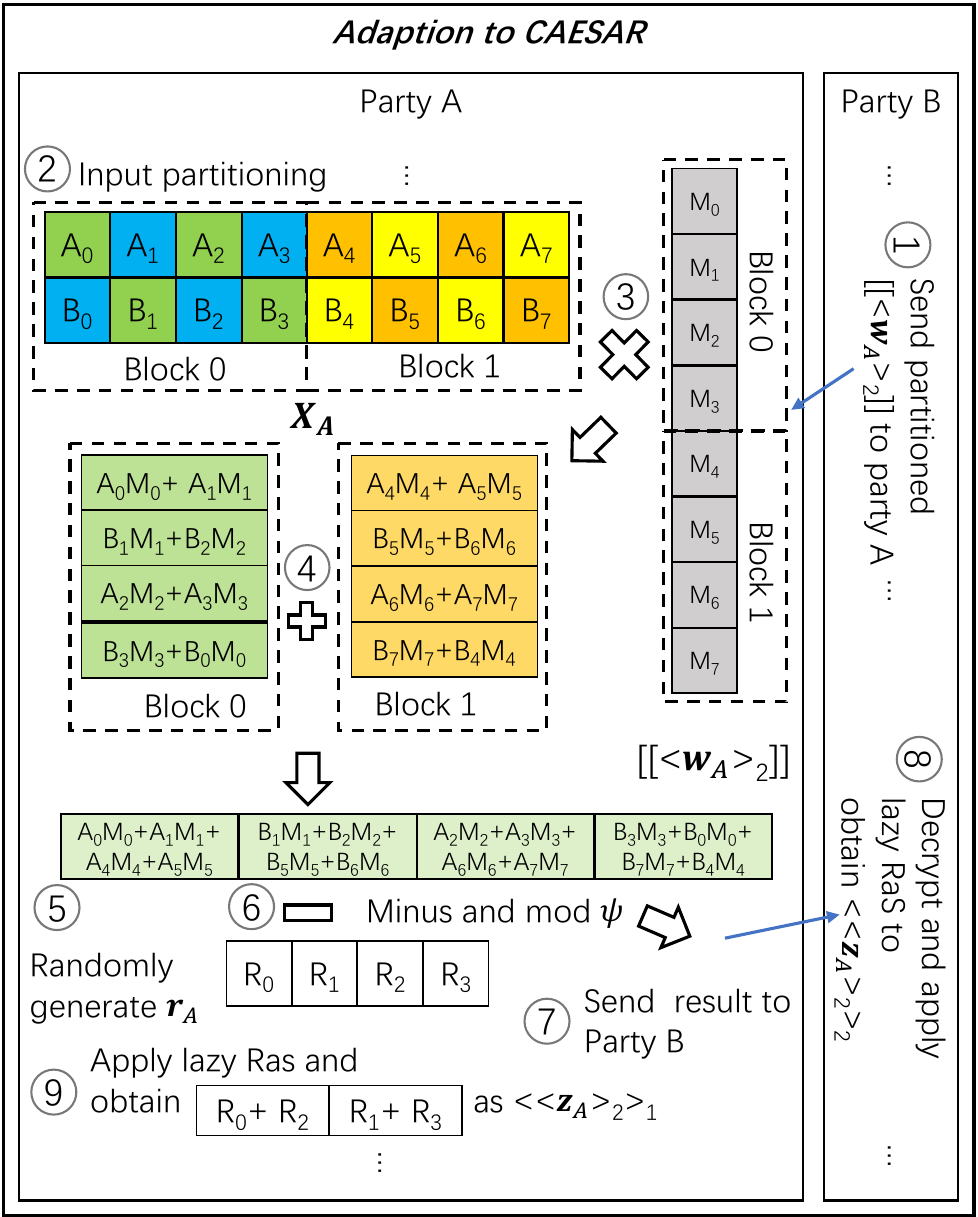}
      \caption{The CAESAR algorithm. We set $m = 8, n = 2, N' = 4$. The input partitioning and lazy RaS techniques are also adopted.}
      \label{fig:caesar_packing}
  \end{subfigure}
  \hfill
  \begin{subfigure}[b]{0.31\textwidth}
      \centering
      \includegraphics[width=\textwidth]{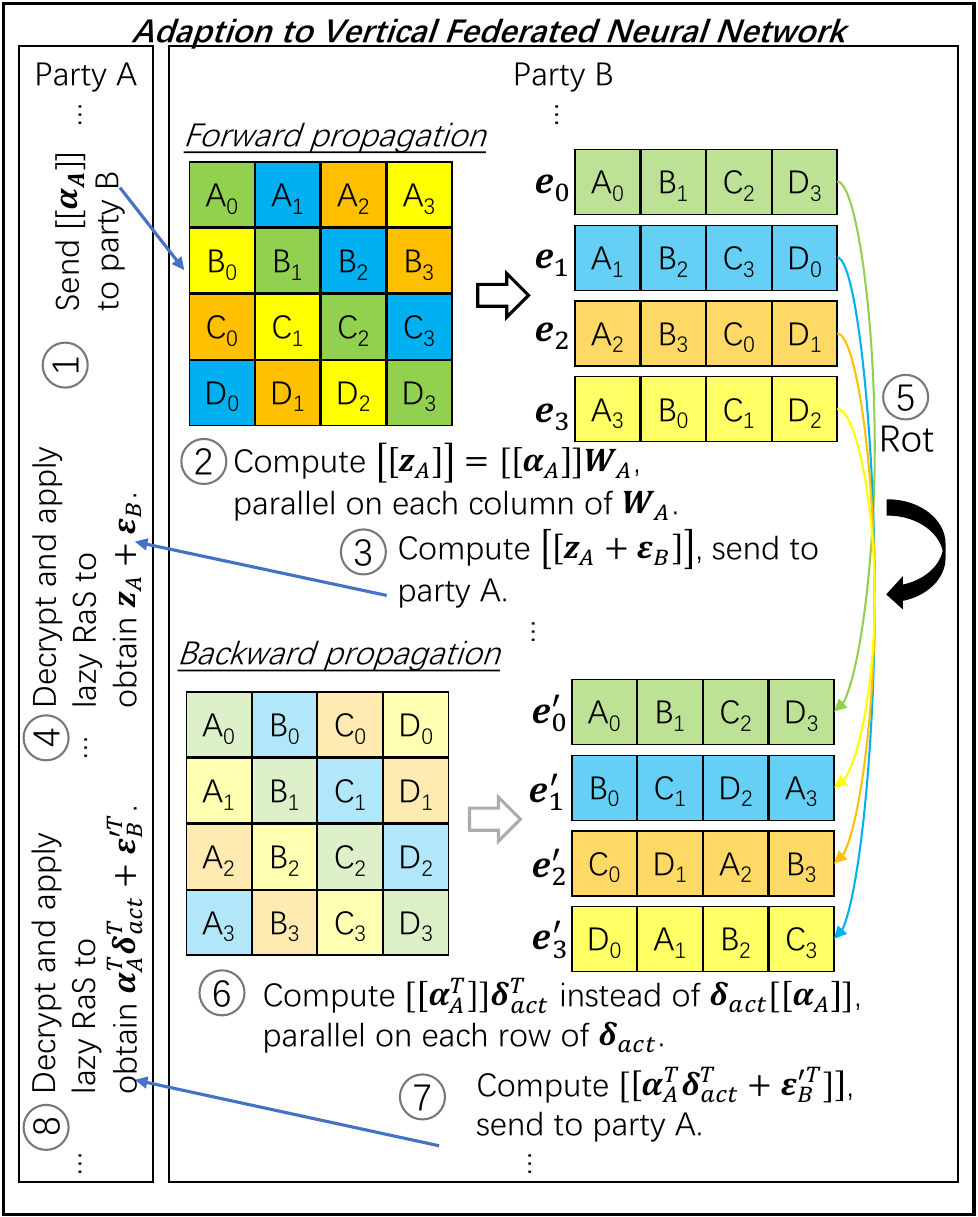}
      \caption{The VFL-NN algorithm. We set $m = n = N' = 4$. The transposed matrices' diagonal conversion is also adopted.}
      \label{fig:hetero_nn_packing}
  \end{subfigure}
     \caption{Illustration of adapting~\sys' MatMult method to three SOTA VFL algorithms.}
     \label{fig:adaption}
\end{figure*}

Considering the~\textbf{C3 (Passive Decryption)} characteristic, we design the Lazy Rotate-and-Sum (RaS) technique. The main idea is to eliminate the remaining ciphertext RaS operations, which contains~\textbf{O3 (Rot)}, and replace them with cleartext sum-up computation. We provide illustrations in both Fig.~\ref{fig:hetero_linr_packing} and Fig.~\ref{fig:caesar_packing}. Taking Fig.~\ref{fig:hetero_linr_packing} as an example, after step 4, $P_A$ obtains $\llbracket [A_0 M_0+A_1M_1,B_1M_1+B_2M_2,A_2M_2+A_3M_3,B_3M_3+B_0M_0,\cdots] \rrbracket$. Naively, a set of time-consuming RaS operations was necessary to aggregate intermediate elements within the ciphertext before sending it to $P_B$ for decryption. However, since $P_B$ performs no additional operations on the ciphertext beyond decryption and directly returns the results, the slots of the ciphertext that need decryption still align correctly with the indices of the resulting cleartext vectors. Consequently, we can defer the RaS operations after the decryption process. Ignoring the masking precedures at step 5\&6, we will explain them in~\S\ref{subsec:vfl_linr}. $P_A$ directly sends $\llbracket [A_0 M_0+A_1M_1,B_1M_1+B_2M_2,A_2M_2+A_3M_3,B_3M_3+B_0M_0,\cdots] \rrbracket$ to $P_B$ and obtains $[A_0 M_0+A_1M_1,B_1M_1+B_2M_2,A_2M_2+A_3M_3,B_3M_3+B_0M_0,\cdots]$ after decryption. Finally, we could substitute the RaS operations with cleartext $A_0 M_0+A_1M_1+A_2M_2+A_3M_3, B_0M_0+B_1M_1+B_2M_2+B_3M_3, C_0M_0+C_1M_1+C_2M_2+C_3M_3, D_0M_0+D_1M_1+D_2M_2+D_3M_3$.

Combining the above three components of our hybrid MatMult method, we can also compute the complexity of~\sys for the cases of $m > N'$ or $n > N'$, shown in Tab.~\ref{tab:complexity_large_m_n}.





\begin{table}[h]
  \centering
  \footnotesize
  \begin{tabular}{c|ccc}
      \toprule
       & $m > N' \geq n$ & $m \leq N' < n$ & $m, n > N'$ \\
      \midrule
      \# O1 (Add) & $\frac{mn-m}{N'}$ & $\frac{nm-N'}{N'}$ & $\frac{mnN'-N'^2}{N'^2}$ \\
      \# O2 (Mult) & $\frac{mn}{N'}$ & $\frac{mn}{N'}$ & $\frac{mn}{N'}$ \\
      \# O3 (Rot) & $0$ & $0$ & $0$ \\
      \# O4 (HstRot) & $n - 1$ & $\frac{mn-n}{N'}$ & $\frac{nN'-n}{N'}$ \\
      $P_B$ to $P_A$ & $1$ RLWE-ct & $\frac{n}{N'}$ RLWE-ct & $\frac{n}{N'}$ RLWE-ct \\
      $P_A$ to $P_B$ & $\frac{m}{N'}$ RLWE-ct & $1$ RLWE-ct & $\frac{m}{N'}$ RLWE-ct \\
      \bottomrule
  \end{tabular}
  \caption{Complexity of~\sys when $m > N'$ or $n > N'$.}
  \label{tab:complexity_large_m_n}
  \vspace{-0.8cm}
\end{table}

\parab{\sys~\textit{v.s.} GALA.} Besides the improvement of diagonal method explained in~\ref{subsec:diagonal},~\sys also has the following advantages over GALA: 1)~\sys's diagonal and input packing method are formally defined with equations. In contrast, GALA introduces its method only using legends and textual descriptions, which could lead to misinterpretation and poor generalizability; 2) GALA lacks mechanisms for input partitioning, whereas~\sys incorporates sophisticated mechanisms that eliminate redundant operations when processing segmented blocks. Our experiments demonstrate a significant improvement in~\sys when the operand size,~\eg, feature dimension or batch size, is large in~\S\ref{sec:experiment}; 3)~\sys extends the concept of lazy RaS in GALA beyond its original application from SS ciphertext to cleartext, thereby increasing its practical utility and relevance across diverse contexts. Additionally, in the VFL scenario,~\sys introduces the innovative mechanism,~\ie, cleartext inverse RaS, to facilitate the implementation of lazy RaS in~\S\ref{subsec:caesar}.

\section{Adaption Design to SOTA VFL Algorithms}
\label{sec:adaption}

In this section, we provide a comprehensive guidance on integrating the proposed PackedHE MatMult method by explaining how to apply it to representative VFL algorithms,~\eg, VFL-LinR~\cite{yang2019federated}, CAESAR~\cite{chen2021homomorphic}, and VFL-NN~\cite{zhang2020additively}. These three algorithms are fundamental and well-recognized by the VFL community~\cite{yang2023survey}. Other algorithms,~\eg,~\cite{huang2022efmvfl,cai2022accelerating,yang2019parallel,yang2019quasi,liu2019communication,li2020efficient,fu2022blindfl}, which also contain MatMult operations, can similarly apply~\sys. Accelerating HE-based VFL algorithms,~\eg,~\cite{cheng2021secureboost}, whose core computation is not MatMult, will be our future work. For CAESAR and VFL-NN, we analyzes their unique algorithmic properties and made three adaption designs to improve efficiency further, as shown in Fig.~\ref{fig:method_overview}:

 
\begin{icompact}
	\item~\textbf{Multiplication Level Reduction}, which optimizes CAESAR's calculation process to reduce the number of accumulated multiplication over a single ciphertext for more efficient PackedHE parameters;	
	\item~\textbf{Cleartext Inverse Rotate-and-Sum (RaS)}, which is designed to make our MatMult method feasible for CAESAR after multiplication level reduction;
 	\item~\textbf{Transposed Matrices' Diagonal Conversion}, which is a conversion mechanism between diagonal encodings of matrix and its transpose to reduce half of the communication cost in VFL-NN.
\end{icompact}

\subsection{Adaption to VFL-LinR}
\label{subsec:vfl_linr}
VFL-LinR~\cite{yang2019federated} securely trains a linear regression model, involving three parties. $P_A,P_B$ hold $\bm{X}_A,\bm{X}_B,\bm{y}_B$ and train weights $\bm{\theta}_A,\bm{\theta}_B$ with the help of a third-party arbiter $P_C$. $P_A, P_B$ exchange encrypted intermediate results $\llbracket \bm{u}_A \rrbracket, \llbracket \bm{d} \rrbracket$ and calculate over them to obtain the encrypted model updates $\llbracket\frac{\delta L}{\delta \bm{\theta}_A}\rrbracket, \llbracket\frac{\delta L}{\delta \bm{\theta}_B}\rrbracket$. Then, $P_A, P_B$ add random masks on the updates and send them to $P_C$ for decryption. Fig.~\ref{fig:hetero_linr_packing} illustrates the computation process of $P_A$, and the operations of $P_B$ could refer to Alg.~\ref{alg:vfl-linr}.




\begin{algorithm}
  \small
  \SetAlgoLined
  \KwIn{$P_B$ receives $\llbracket \bm{u}_A \rrbracket$ from $P_A$ (step 2 in Tab. 1 of VFL-LinR~\cite{yang2019federated})\;}
  \KwOut{$P_A, P_B$ obtain $\frac{\delta L}{\delta \bm{\theta}_A}, \frac{\delta L}{\delta \bm{\theta}_B}$, respectively (step 4 in Tab. 1 of VFL-LinR)\;}

  $P_B$ computes replicated $\llbracket \bm{d} \rrbracket = \llbracket \bm{u}_A \rrbracket + (\bm{u}_B - \bm{y})$ and sends it to $P_A$\;
  $P_A, P_B$ conduct vertical input packing over plaintext matrix $\bm{X}_A^T, \bm{X}_B^T$, respectively\;
  
  \While{$ i=0,1,...,\frac{mn}{N'}-1$}{
    $P_A, P_B$ generate plaintext $\tilde{e}_{A,i}, \tilde{e}_{B,i}$ with diagonal encodings, respectively\;
    $P_A, P_B$ conduct multiplication $\tilde{e}_{A,i} \times \mbox{HstRotL}(\llbracket \bm{d} \rrbracket,i), \tilde{e}_{B,i} \times \mbox{HstRotL}(\llbracket \bm{d} \rrbracket, i)$\;
  }
  $P_A, P_B$ conduct addition over the above $\frac{mn}{N'}$ resulted ciphertexts separately\;
  $P_A, P_B$ generate the random mask $\bm{r}_A, \bm{r}_B$ and add the mask on the result of the previous step\;
  $P_A, P_B$ send the masked ciphertext to $P_C$\;
  $P_C$ decrypts the received messages and sends the cleartexts back to $P_A, P_B$, respectively\;
  $P_A, P_B$ remove $\bm{r}_A, \bm{r}_B$ and apply lazy RaS to obtain the gradients $\frac{\delta L}{\delta \bm{\theta}_A}, \frac{\delta L}{\delta \bm{\theta}_B}$.
  
  \caption{Modification to VFL-LinR Algorithm}
  \label{alg:vfl-linr}
\end{algorithm}

Starting from step 2 in Tab. 1 of VFL-LinR~\cite{yang2019federated}, $P_B$ receives the $\llbracket \bm{u}_A \rrbracket$ from $P_A$, computes the $\llbracket \bm{d} \rrbracket$ with the replicated encoding, and sends it to $P_A$. Then, $P_A, P_B$ conduct vertical input packing over plaintext matrix $\bm{X}_A^T, \bm{X}_B^T$, and generate $\frac{mn}{N'}$ plaintext $\bm{e}_{A,i}, \bm{e}_{B,i}$. For each plaintext, $P_A, P_B$ conduct multiplication with corresponding the rotated ciphertext $\mbox{HstRotL}(\llbracket \bm{d} \rrbracket, i)$. Next, $P_A, P_B$ separately sum up the results. After that, $P_A, P_B$ send the masked ciphertexts to $P_C$. $P_C$ decrypts them and sends the results back. $P_A, P_B$ remove the masks and apply the lazy RaS to obtain gradients for model update. The process ends at step 4 in Tab. 1 of VFL-LinR.




\subsection{Adaption to CAESAR}
\label{subsec:caesar}
CAESAR~\cite{chen2021homomorphic} is a vertical federated logistic regression algorithm combining HE and SS to enhance the security, containing two parties. $P_A,P_B$ distributely train on $\bm{X}_A,\bm{X}_B, \bm{y}_B$ to obtain weights $\bm{w}_A, \bm{w}_B$. CAESAR is a SOTA end-to-end training algorithm. Our modifications are mainly located in Protocol 1 of CAESAR. Besides, we also make extra designs,~\ie,~\textbf{multiplication level reduction} and~\textbf{plaintext inverse RaS}, to further improve efficiency. 

Taking the prediction calculation process as an example, we show the modification to Protocol 1 of CAESAR. $P_A, P_B$ initialize $\bm{w}_A, \bm{w}_B$, generate and exchange shares of model weights. More specifically, $P_A$ generates shares $\langle \bm{w}_A \rangle _1, \langle \bm{w}_A \rangle_2$ ($\langle \bm{w}_A \rangle_1 + \langle \bm{w}_A \rangle_2 = \bm{w}_A$), holds $\langle \bm{w}_A \rangle_1$, and sends $\langle \bm{w}_A \rangle_2$ to $P_B$, where $\langle \cdot \rangle$ stands for a secret share. Next, $P_A$ conducts $\bm{X}_A \llbracket \langle \bm{w}_A \rangle_2 \rrbracket$ at line 11 in Algorithm 1 of CAESAR~\cite{chen2021homomorphic}.


\begin{algorithm}
  \small
  \SetAlgoLined
  \KwIn{$P_A$ receives partitioned $\llbracket \langle \bm{w}_A \rangle_2 \rrbracket$ from $P_B$ (line 1 in Protocol 1 of CAESAR~\cite{chen2021homomorphic})\;}
  \KwOut{$P_A, P_B$ get $\langle \langle\bm{z}_A \rangle_2 \rangle_1, \langle \langle\bm{z}_A \rangle_2 \rangle_2$, respectively (output in Protocol 1 of CAESAR)\;}
  $P_A$ conducts input partitioning on matrix $\bm{X}_A$ and obtains $\frac{m}{N'}$ matrix blocks\;
  \While{$ j=0,1,...,\frac{m}{N'}$}{
    $P_A$ performs~\sys-MatMult in block $j$ without the final RaS operations\;
  }
  $P_A$ adds up the intermediate results of $\frac{m}{N'}$ blocks\;
  $P_A$ randomly generates vector $\bm{r}_A$\;
  $P_A$ subtracts the sum of intermediate results with $\bm{r}_A$, following Protocol 2 in CAESAR\;
  $P_A$ sends the subtraction result to $P_B$\;
  $P_B$ decrypts the received ciphertext and applies lazy RaS to obtain $\langle \langle\bm{z}_A \rangle_2 \rangle_2$\;
  $P_A$ conducts cleartext RaS over $\bm{r}_A$ and gets $\langle \langle\bm{z}_A \rangle_2 \rangle_1$.
  \caption{Modification to CAESAR Algorithm}
  \label{alg:caesar}
\end{algorithm}

As shown in Fig.~\ref{fig:caesar_packing} and Alg.~\ref{alg:caesar}, at first, $P_B$ divides $\langle \bm{w}_A \rangle_2$ into $\frac{m}{N'}$ blocks, encrypts, and sends them to $P_A$. $P_A$ conducts input partitioning on matrix $\bm{X}_A$ and obtains blocks. For each vector and matrix block, $P_A$ performs the MatMult method without the final RaS operations. Next, $P_A$ conducts addition over the results of all blocks, randomly generates vector $\bm{r}_A$ and substracts it from the resulting ciphertext. Finally, $P_A$ sends the outcome to $P_B$. $P_B$ decrypts it and applies lazy RaS. $P_A$ conducts cleartext RaS over $\bm{r}_A$. 


\subsubsection{Multiplication Level Reduction.}
The efficiency of PackedHE is highly related to its parameters. Parameters allowing less multiplication refer to more efficient PackedHE operations. Therefore, we modify the computation process of CAESAR to reduce the multiplication level. More detailedly, in line 21 of the Algorithm 1 in CAESAR, $\llbracket \bm{e} \rrbracket^T = \llbracket \hat{\bm{y}} \rrbracket - \bm{y} = q_0 + q_1 \llbracket \bm{z} \rrbracket + q_2 \llbracket \bm{z}^3 \rrbracket - \bm{y}$ is computed, which consumes one multiplication level,~\ie, $q_1 \llbracket \bm{z} \rrbracket$ and $q_2 \llbracket \bm{z}^3 \rrbracket$. Then, $\llbracket \bm{g}_B \rrbracket = \llbracket \bm{e} \rrbracket^T \bm{X}_B$ is computed based on the resulted $\llbracket \bm{e} \rrbracket^T$, which needs one more multiplication level. We integrate them to:


\begin{equation}
\label{eq:caesar}
\begin{aligned}
	\llbracket \bm{g}_B \rrbracket & = \llbracket e \rrbracket^T \bm{X}_B = (\llbracket \hat{\bm{y}} \rrbracket - \bm{y})^T \bm{X}_B = (q_0 + q_1 \llbracket \bm{z} \rrbracket + q_2 \llbracket \bm{z}^3 \rrbracket - \bm{y}) \bm{X}_B \\
	& = (q_0 \overrightarrow{\bm{1}} - \bm{y}) \bm{X}_B + (q_1 \bm{X}_B) \llbracket \bm{z} \rrbracket + (q_2 \bm{X}_B) \llbracket \bm{z}^3 \rrbracket,
\end{aligned}
\end{equation}
which decreases the multiplication level by one since the computation of $(q_1 \bm{X}_B) \llbracket \bm{z} \rrbracket$ and $(q_2 \bm{X}_B) \llbracket \bm{z}^3 \rrbracket$ only need one multiplication level.

\subsubsection{Cleartext Inverse RaS.} 
However, Eq.~\ref{eq:caesar} introduces extra cleartext addition term $(q_0 \overrightarrow{\bm{1}} - \bm{y}) \cdot{X}_B$ before secretly sharing $\llbracket \bm{g}_B \rrbracket$. We cannot directly perform lazy RaS over the shares because the addition term needs no inner sum operation. One option is to separately share the above cleartext addition term and the remaining ciphertext term, only conducting lazy RaS over the share of the ciphertext term. But this approach increases communication cost. Therefore, we design the cleartext inverse RaS method, which iteratively conducts split and concatenation over the cleartext plaintext term before addition, which properly aligns the number of required RaS between cleartext and ciphertext addition terms. For example, after one inverse RaS, cleartext vector $[M_0, M_1]$ turns to $[M_{0,0}, M_{1,0}, M_{0,1}, M_{1,1}]$, where $M_0 = M_{0,0} + M_{0,1}, M_1 = M_{1,0} + M_{1,1}$. 




\subsection{Adaption to VFL-NN}
\label{subsec:vfl_nn}

VFL-NN~\cite{zhang2020additively} implements a distributed neural network~\cite{abiodun2018state} via the SplitNN~\cite{vepakomma2018split} architecture. It divides NN layers into $P_A, P_B$, separately owning $\bm{X}_A,\bm{X}_B, \bm{y}_B$. VFL-NN, capable of handling more complex networks, addresses "harder" tasks. We also refer to the VFL-NN implementation in FATE~\cite{liu2021fate}. In VFL-NN, $P_A$ feeds data into the bottom layer and obtains embedding $\bm{\alpha}_A \in \mathbb{R}^{m \times n}$, where $m$ is the batch size, and $n$ is the number of neurons in the bottom layer. Then, $P_A$ encrypts $\bm{\alpha}_A$ into $\llbracket \bm{\alpha}_A \rrbracket$ and sends them to $P_B$ who contains the interactive and top layers. Next, $P_B$ conducts two separate MatMult $\llbracket \bm{\alpha}_A \rrbracket \cdot \bm{W}_A$ and $\bm{\delta}_{\mbox{act}} \cdot \llbracket \bm{\alpha}_A \rrbracket = (\llbracket \bm{\alpha}_A^T \rrbracket \cdot \bm{\delta}_{\mbox{act}}^T)^T$ in the forward and backward propagation calculation. We have two observations: 1) the MatMult is between matrices; 2) $\llbracket \bm{\alpha}_A \rrbracket$ locates at opposite positions in two MatMult. For the first observation, we choose to apply diagonal encoding on the ciphertext matrix instead of the cleartext matrix to eliminate all the existing~\textbf{O4 (HstRot)}, which is unprecedented in earlier works. 

\begin{algorithm}
  \small
  \SetAlgoLined
  \KwIn{$P_B$ receives diagonally encoded $\llbracket \bm{\alpha}_A \rrbracket$ from $P_A$ (line 5 in Alg. 1 of VFL-NN~\cite{zhang2020additively})\;}
  \KwOut{$P_A$ obtains $\bm{z}_A + \bm{\epsilon}_B$ (line 8 in Alg. 1 of VFL-NN)\;}

  $P_B$ randomly generates $\bm{\epsilon}_B \in \mathbb{Z}^{m \times n'}$\;
  \While{$k=0,1,...,n'$}{
    $P_B$ conducts~\sys-MatMult $\llbracket \bm{z}_A[:,k] \rrbracket = \llbracket \bm{\alpha}_A \rrbracket \bm{W}_A[:,k]$\;
    $P_B$ computes $\llbracket \bm{z}_A[:,k] \rrbracket + \bm{\epsilon}_B[:,k]$\;
  }

  $P_B$ obtains $\llbracket \bm{z}_A + \bm{\epsilon}_B \rrbracket = \{ \llbracket \bm{z}_A[:,k] \rrbracket + \bm{\epsilon}_B[:,k] \}_{k=0,1,...,n'}$ and sends it to $P_A$\;
  $P_A$ decrypts the received ciphertexts and applies lazy RaS to obtain $\bm{z}_A + \bm{\epsilon}_B$.
  
  \caption{Modification to the Forward Propagation Process of VFL-NN Algorithm }
  \label{alg:vfl-nn-forward}
\end{algorithm}

\subsubsection{Transposed Matrices' Diagonal Conversion.}
For the second observation, we designed the~\textbf{transposed matrices' diagonal conversion} mechanism, which further reduces half of the communication cost. Therefore, instead of transmitting both $\llbracket \bm{\alpha}_A \rrbracket$ and $\llbracket \bm{\alpha}_A^T \rrbracket$ from $P_A$ to $P_B$, we could only transmit $\llbracket \bm{\alpha}_A \rrbracket$. The mechanism allows conversion between diagonal encodings $\tilde{e}$ and $\tilde{e}'$ of transposed matrices:

\begin{equation}
\label{eq:conversion}
\begin{aligned}
	\tilde{e}_i' = \mbox{RotR}( & \tilde{e} [(\min (m, n) - i) \hspace{-0.2cm} \mod \min (m, n)],\\
	& \max (m, n) - \min (m, n) + i).
\end{aligned}
\end{equation}





Shown in Alg.~\ref{alg:vfl-nn-forward} and Fig.~\ref{fig:hetero_nn_packing}, our modification to forward propagation process starts from line 5 in Alg. 1 of VFL-NN~\cite{zhang2020additively}. First, $P_A$ diagonally encodes, encrypts $\llbracket \bm{\alpha}_A \rrbracket$, and sends it to $P_B$. $P_B$ generates random mask $\bm{\epsilon}_B \in \mathbb{R}^{m \times n'}$, where $n'$ is the number of neurons in the interactive layer $\bm{W}_A \in \mathbb{R}^{n \times n'}$. For each column $k$ of $\bm{W}_A$ in parallel, $P_B$ conducts MatMult between $\llbracket \bm{\alpha}_A \rrbracket$ and $\bm{W}_A$ to obtain $\llbracket \bm{z}_A[:k] \rrbracket$, where $[:,k]$ represents the $k$ column of matrix. The MatMult is efficient since we eliminate all~\textbf{O4 (HstRot)}. Besides, $P_B$ masks each resulted ciphertext. Next, $P_B$ sends them to $P_A$ for decryption and lazy RaS. In the idle period of waiting for $P_A$'s response, $P_B$ conducts~\textbf{transposed matrices' diagonal conversion} according to Eq.~\ref{eq:conversion} and obtains $\llbracket \bm{\alpha}_A^T \rrbracket$. 



\begin{algorithm}
  \small
  \SetAlgoLined
  \KwIn{$P_B$ obtains $\llbracket \bm{\alpha}_A^T \rrbracket$ via Eq.~\ref{eq:conversion} (line 3 in Alg. 2 of VFL-NN~\cite{zhang2020additively})\;}
  \KwOut{$P_A$ obtains $\bm{\delta}_{\mbox{act}} \bm{\alpha}_A + \bm{\epsilon}'_B$ (line 7 in Alg. 2 of VFL-NN)\;}

  $P_B$ randomly generates $\bm{\epsilon}'_B \in \mathbb{Z}^{n' \times n}$\;
  \While{$k=0,1,...,n'$}{
    $P_B$ conducts~\sys-MatMult $\llbracket \bm{\alpha}_A^T \bm{\delta}_{\mbox{act}}^T[:,k] \rrbracket = \llbracket \bm{\alpha}_A^T \rrbracket \bm{\delta}_{\mbox{act}}^T[:,k]$\;
    $P_B$ computes $\llbracket \bm{\alpha}_A^T \bm{\delta}_{\mbox{act}}^T[:,k] \rrbracket + \bm{\epsilon}_B^{'T}[:,k]$\;
  }
  
  $P_B$ obtains $\llbracket \bm{\alpha}_A^T \bm{\delta}_{\mbox{act}}^T + \bm{\epsilon}_B^{'T} \rrbracket = \{ \llbracket \bm{\alpha}_A^T \bm{\delta}_{\mbox{act}}^T[:,k] \rrbracket + \bm{\epsilon}_B^{'T}[:,k] \}_{k=0,1,...,n'}$ and sends it to $P_A$\;
  $P_A$ decrypts the received ciphertexts, applies lazy RaS, and conducts transpose operation on them to obtain $\bm{\delta}_{\mbox{act}} \bm{\alpha}_A + \bm{\epsilon}'_B = (\bm{\alpha}_A^T \bm{\delta}_{\mbox{act}}^T + \bm{\epsilon}_B^{'T})^T$.
  
  \caption{Modification to the Backward Propagation Process of VFL-NN Algorithm}
  \label{alg:vfl-nn-backward}
\end{algorithm}


Shown in Alg.~\ref{alg:vfl-nn-backward} and Fig.~\ref{fig:hetero_nn_packing}, our modification to the backward propagation process starts from line 3 in Alg. 2 of VFL-NN~\cite{zhang2020additively}. First, $P_B$ randomly generates mask $\bm{\epsilon}'_B \in \mathbb{Z}^{n' \times n}$. Then, for each column $k$ of $\bm{\delta}_{\mbox{act}}^T$, $P_B$ conducts MatMult between $\llbracket \bm{\alpha}_A^T \rrbracket$ and $\bm{\delta}_{\mbox{act}}^T[:,k]$, where $\bm{\delta}_{\mbox{act}} \in \mathbb{R}^{n' \times m}$ is the error back-propagated from the top layer. Besides, $P_B$ conducts addition between $\llbracket \bm{\alpha}_A^T \rrbracket + \bm{\delta}_{\mbox{act}}^T[:,k]$ and $\bm{\epsilon}_B^{'T}[:,k]$. Finally, $P_B$ collects $\llbracket \bm{\alpha}_A^T \bm{\delta}_{\mbox{act}}^T + \bm{\epsilon}_B^{'T} \rrbracket$ and sends it to $P_A$. $P_A$ decrypts them, conducts lazy RaS and transpose to obtain $\bm{\delta}_{\mbox{act}} \bm{\alpha}_A + \bm{\epsilon}'_B$.

\section{Correctness and Security Analysis}
\label{sec:security}
\sys innovates on the top of but didn't change basic PackedHE operations. Therefore, ciphertext correctness/security is fully protected by PackedHE, which is well recognized~\cite{juvekar2018gazelle,zhang2021gala}. Besides,~\sys substitutes Paillier with PackedHE in HE-based VFL for cryptographic computation, which enhances the ciphertext security since PackedHE is quantum-resilient~\cite{nejatollahi2019post} and Paillier is vulnerable to post-quantum attacks~\cite{shor1999polynomial}.

We empirically show that our end-to-end training accuracy loss is small enough in~\S\ref{subsec:exp_e2e}. In addition, security of VFL protocols varies based on their origionally designed exposure of cleartexts during training. For instance, VFL-LinR and VFL-NN are more susceptible to attacks than CAESAR due to greater information exposure. CAESAR exposes no intermediate cleartext results during training. As a plug-in method,~\sys also didn't modify the original federated protocols~\cite{yang2019federated,chen2021homomorphic,zhang2020additively}. Thus, their security analysis of VFL protocols, under various threat models,~\eg, semi-honest/malicious, remains valid and can be our reference.

\begin{figure*}[h]
    \centering
    \begin{subfigure}[b]{0.33\textwidth}
           \centering
           \includegraphics[width=\textwidth]{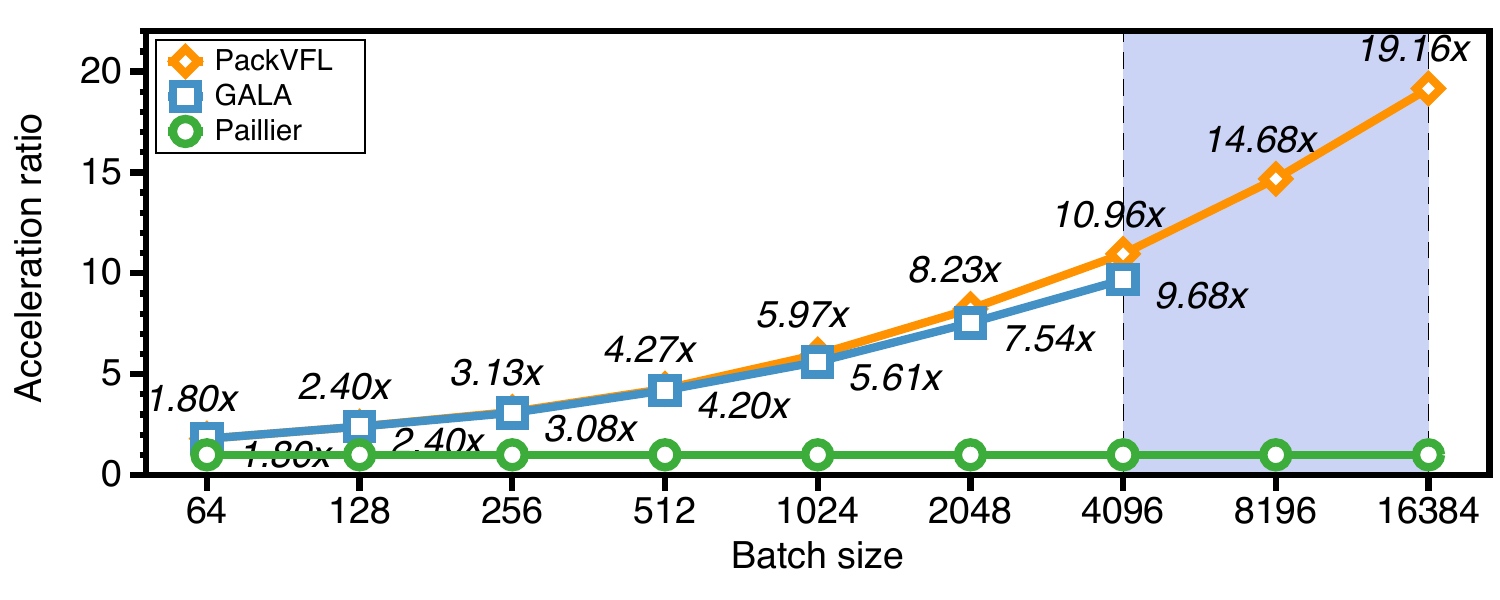}
           \caption{VFL-LinR with varying batch sizes.}
       \label{fig:experiment_linr_batch}
       \end{subfigure}
       \hfill
       \begin{subfigure}[b]{0.33\textwidth}
           \centering
           \includegraphics[width=\textwidth]{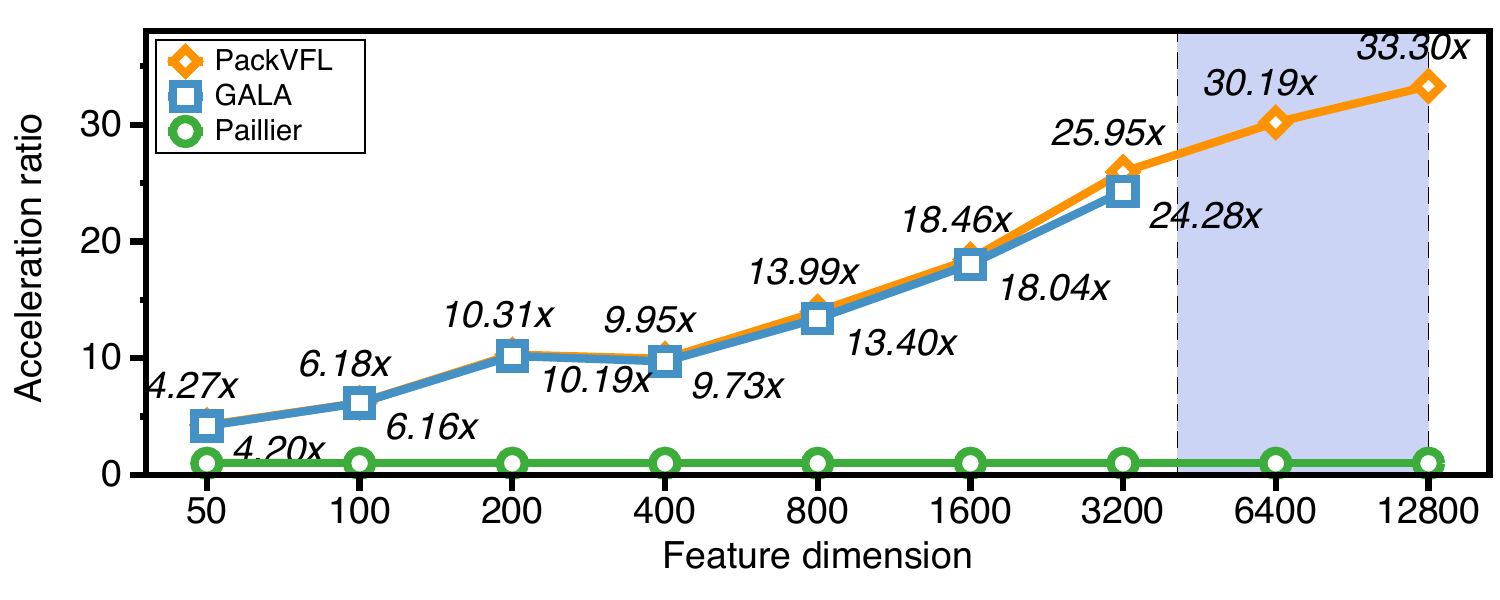}
           \caption{VFL-LinR with varying feature dimensions.}
       \label{fig:experiment_linr_feat}
       \end{subfigure}
       \hfill
    \begin{subfigure}[b]{0.33\textwidth}
           \centering
           \includegraphics[width=\textwidth]{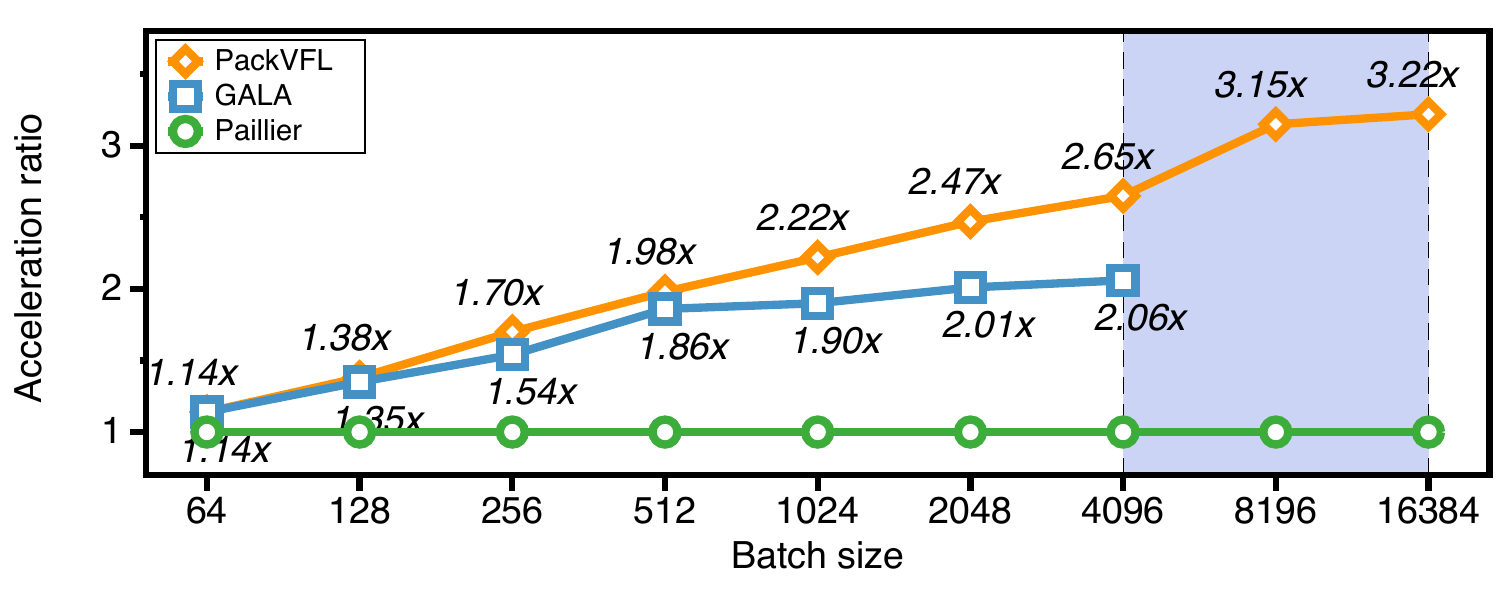}
           \caption{CAESAR with varying batch sizes.}
       \label{fig:experiment_caesar_batch}
       \end{subfigure}
       \hfill
       \begin{subfigure}[b]{0.33\textwidth}
           \centering
           \includegraphics[width=\textwidth]{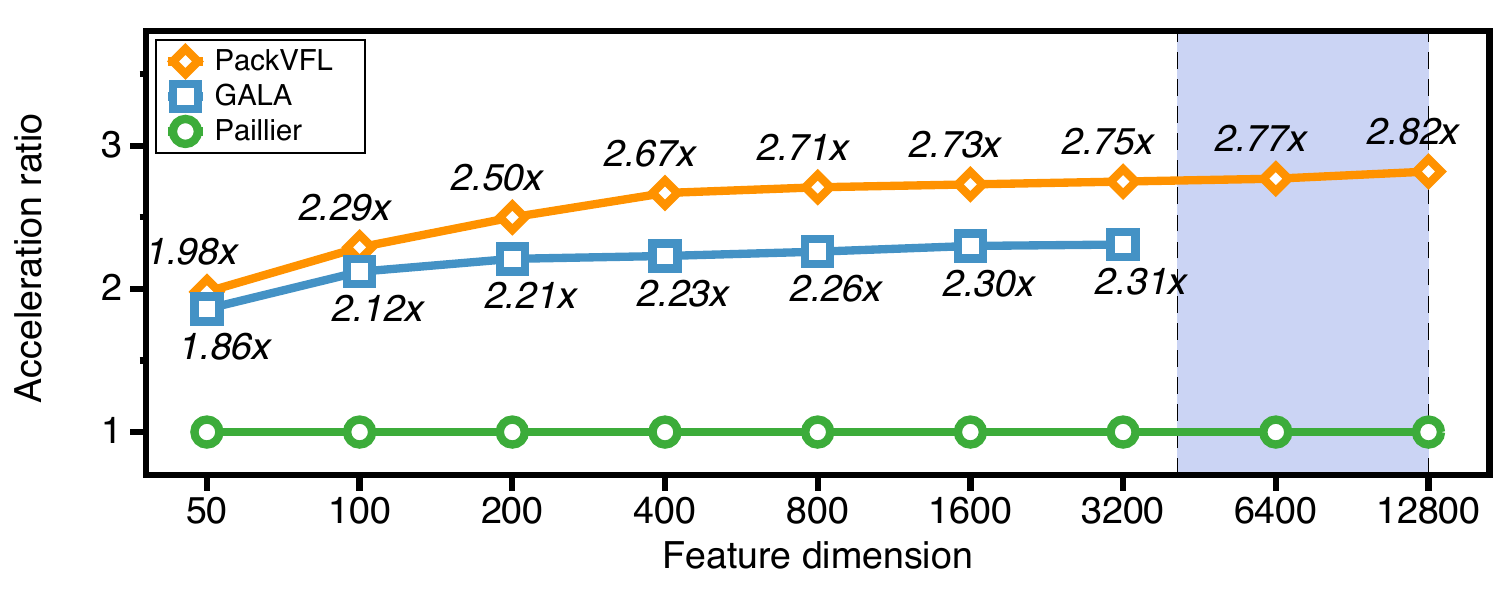}
           \caption{CAESAR with varying feature dimensions.}
       \label{fig:experiment_caesar_feat}
       \end{subfigure}
       \hfill
       \begin{subfigure}[b]{0.33\textwidth}
           \centering
           \includegraphics[width=\textwidth]{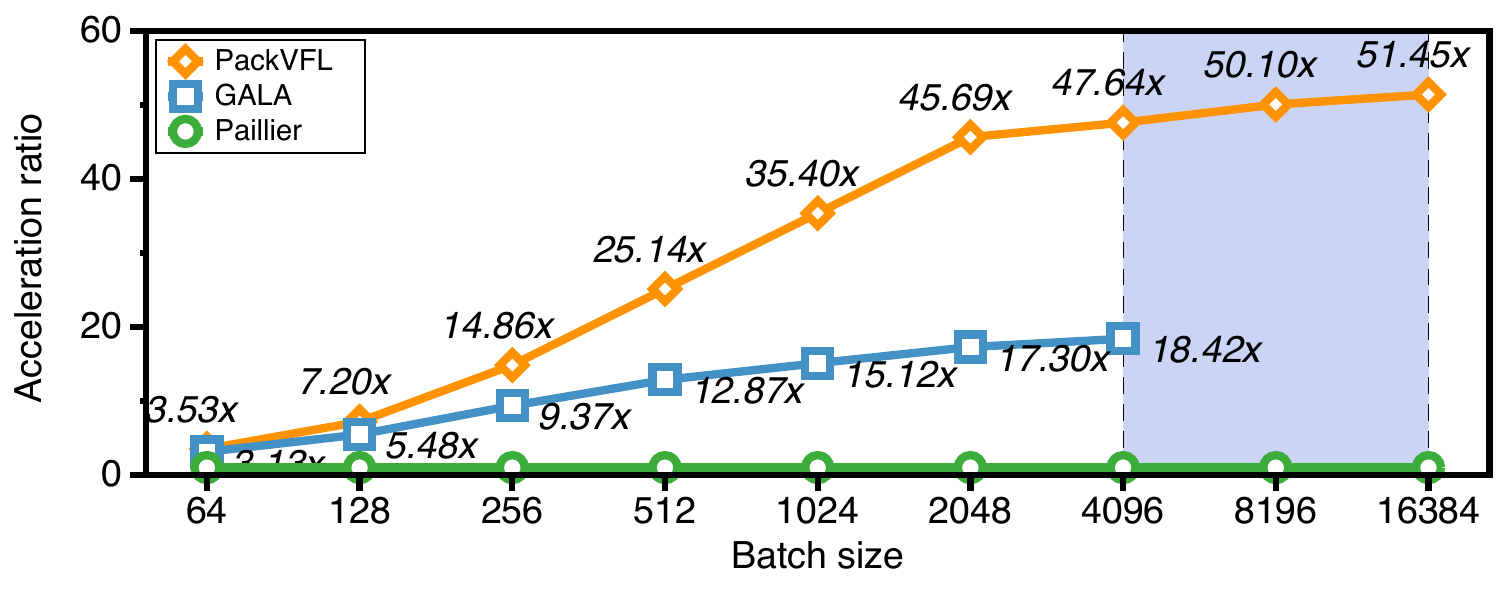}
           \caption{VFL-NN with varying batch sizes.}
       \label{fig:experiment_hetero_nn_batch}
       \end{subfigure}
       \hfill
       \begin{subfigure}[b]{0.33\textwidth}
           \centering
           \includegraphics[width=\textwidth]{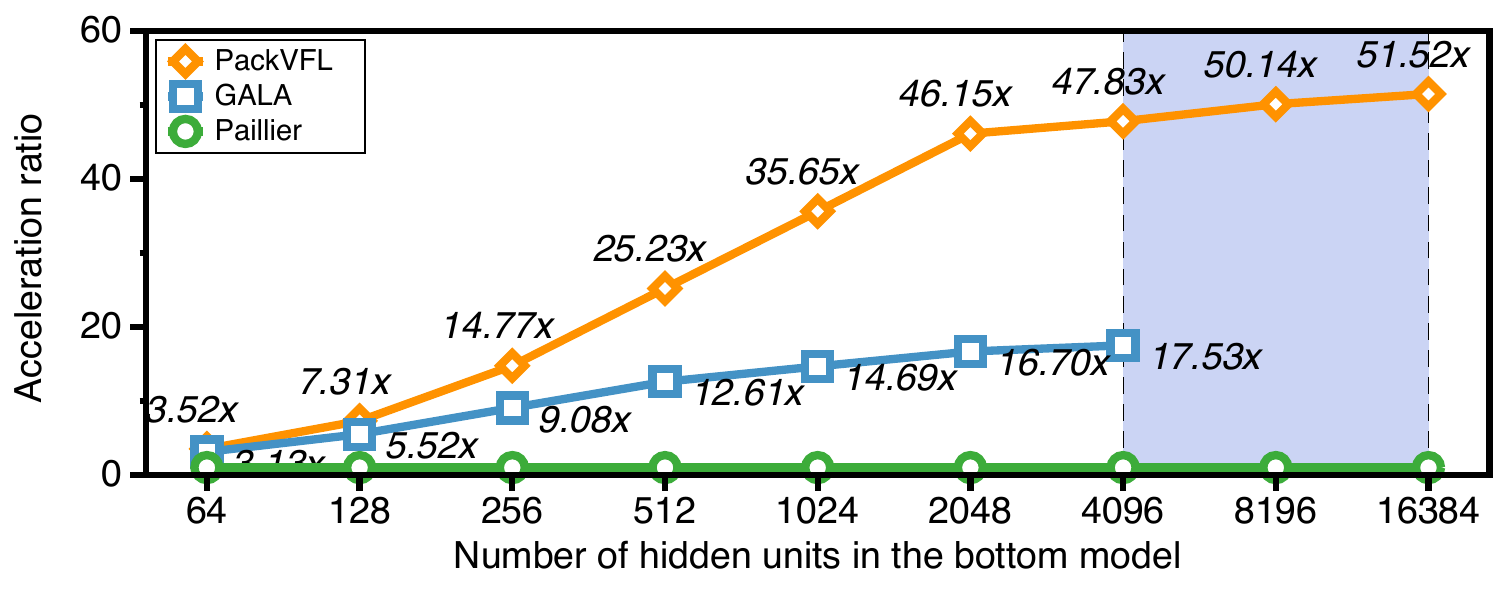}
           \caption{VFL-NN with varying hidden units.}
       \label{fig:experiment_hetero_nn_feat}
       \end{subfigure}
    \caption{End-to-end acceleration of~\sys and GALA for original Paillier-based VFL-LinR, CAESAR and VFL-NN algorithms. The ratio is computed using the time of each training epoch.}
    \label{fig:e2e_experiment}
  \end{figure*}

\section{Experiment}
\label{sec:experiment}

\parab{Setup and Implementation.} The experiments are conducted on three x86 servers, each with 128GB memory and 40-core Intel Xeon Glod 5115 CPU. The bandwidth and latency among servers are 50MB/s and 20ms, respectively. We build our framework based on FATE~\cite{liu2021fate} and Lattigo~\cite{lattigo}. We compile the CKKS~\cite{cheon2017homomorphic} interfaces of Lattigo with CGO~\footnote{https://golang.org/cmd/cgo/}, implement our MatMult method over the compiled interfaces, and integrate it into VFL-LinR, CAESAR, and VFL-NN of FATE. The implementation of GALA refers to the open-source codes~\footnote{https://github.com/mc2-project/delphi}.


\parab{Dataset.} We utilize the SUSY dataset~\cite{baldi2014searching} to conduct the efficiency experiments, focusing on both computation and communication aspects. To meet different requirements of data size, we adjust the sample count and feature dimensions through sampling. Additionally, we use the NUS-WIDE dataset~\cite{nus-wide-civr09} for the accuracy experiments, since the NUS-WIDE dataset contains heterogeneous features,~\ie, text and image features, and is naturally apt for the VFL scenario~\footnote{We didn't use other datasets with homogeneous features, such as MNIST~\cite{lecun1998gradient}, CIFAR~\cite{krizhevsky2009learning}, and LEAF~\cite{caldas2018leaf}.}. We separate its text and image features into two parties and create the vertically partitioned data situation. From the 81 groundtruth concepts, we select the most balanced one to serve as the label for our binary classification task.


\parab{Baseline.} The baselines of the end-to-end experiment are utilizing Paillier and GALA~\cite{zhang2021gala} for MatMult computation in VFL-LinR, CAESAR, and VFL-NN, respectively. When batch size and feature dimension exceed $N' = N/2 = 4096$, we do not show the results of GALA because GALA have no specific design of input partitioning. 

 


\parab{Selection of PackedHE Parameters.}
The parameters of PackedHE,~\ie, $N,q$, are pretty relative to the efficiency and security. $q$ also decides the multiplication level. We set $N = 8192$ for all the experiments. We set $\log q = 156$ for the naive method and GALA in CAESAR, which allows two multiplication operations, and sets $\log q = 122$ for the other methods, which allows one multiplication operation. According to~\cite{albrecht2021homomorphic}, the involved PackedHE MatMult methods offer a 192-bit security level. Therefore, we set the security level of the adopted Paillier as 128-bit~\cite{catalano2001bit} for a fair comparison.

\subsection{End-to-End VFL}
\label{subsec:exp_e2e}

We show the end-to-end acceleration of GALA and~\sys over Paillier for the VFL training process in Fig.~\ref{fig:e2e_experiment}. We could conclude that~\sys always has the most significant acceleration ratio, which reflects speed gains over Paillier-based VFL algorithms, in different situations. Additionally, we demonstrate that~\sys does not compromise the accuracy of the VFL training process in Tab.~\ref{tab:accuracy}.

\subsubsection{Accelerating VFL-LinR.}
In Fig.~\ref{fig:experiment_linr_batch} and Fig.~\ref{fig:experiment_linr_feat}, we vary the batch size and feature dimension, fixing the feature dimension and batch size as 50 and 512, respectively. As batch size and feature dimension increase, the speedups of GALA and~\sys over Paillier also increase.~\sys is more efficient than GALA from the beginning, with fewer rotation operations~\textbf{O3 (Rot)} and~\textbf{O4 (HstRot)}. \sys achieves the largest acceleration ratio up to $33.30\times$ over Paillier-based VFL-LinR when feature dimension extends to 12800 in Fig.~\ref{fig:experiment_linr_feat}. As explained in~\S\ref{subsec:input_pp}, scenarios involving large feature dimensions are quite typical in VFL.


\subsubsection{Accelerating CAESAR.}
In Fig.~\ref{fig:experiment_caesar_batch} and Fig.~\ref{fig:experiment_caesar_feat}, we also fix feature dimension and batch size as 50 and 512, separately. The speedups of GALA and~\sys over Paillier in CAESAR are all smaller than those in VFL-LinR because CAESAR also contains many SS operations, which~\sys cannot accelerate. Besides, the performance of GALA degrades more than~\sys because they have no design for reducing multiplication level, thus leading to a less efficient PackedHE parameter $\log q = 156$.~\sys still has the best performance from the beginning, with the largest speedup over Paillier up to $3.22\times$.





\begin{table*}[h]
  \footnotesize
	\centering
	\begin{tabular}{ccc|ccc|ccc}
		\toprule
		\multirow{2}{*}{Squared loss} & \multicolumn{2}{c}{\underline{VFL-LinR}} & \multirow{2}{*}{Log loss} & \multicolumn{2}{c}{\underline{CAESAR}} & \multirow{2}{*}{Log loss} & \multicolumn{2}{c}{\underline{VFL-NN}} \\
		& Paillier & \sys & & Paillier & \sys & & Paillier & \sys \\
		\midrule
		epoch $ = 1$ & 1.4375 & 1.4283 (\textbf{\underline{-0.0092}}) & epoch $ = 1$ & 0.5563 & 0.5526 (-0.0037) & epoch $ = 1$ & 0.6442 & 0.6515 (\textbf{\underline{+0.0073}}) \\
		epoch $ = 5$ & 0.6991 & 0.6918 (-0.0073) & epoch $ = 5$ & 0.3287 &  0.3329 (+0.0042) & epoch $ = 5$ & 0.3797 & 0.3708 (-0.0088) \\
		epoch $ = 10$ & 0.4643 & 0.4661 (+0.0018) & epoch $ = 10$ & 0.2748 & 0.2691 (-0.0057) & epoch $ = 10$ & 0.2144 & 0.2173 (+0.0029) \\
		\midrule
		AUC & 0.9025 & 0.9038 (+0.0013) & AUC & 0.9652 & 0.9717 (+0.0065) & AUC & 0.9815 & 0.9783 (-0.0032) \\
		\bottomrule
	\end{tabular}
	\caption{Accuracy comparison between Paillier-based VFL algorithms and~\sys. We configured identical hyperparameters for both, such as setting the learning rate to 0.001. We present an analysis that includes both the intermediate loss and the final Area Under the Curve (AUC) score on the training samples, with differences noted in parentheses.}
	\label{tab:accuracy}
\end{table*}

\subsubsection{Accelerating VFL-NN.}
In Fig.~\ref{fig:experiment_hetero_nn_batch} and Fig.~\ref{fig:experiment_hetero_nn_feat}, we vary the batch size and number of hidden units in the bottom model, fixing the number of hidden units in the bottom model and batch size as 32 and 32, respectively. Besides, the feature dimension is 50, the number of hidden units in the interactive layer is 32, and the top model is linear. We show that~\sys's acceleration for VFL-NN is much more significant than acceleration for both VFL-LinR and CAESAR. The reason is that~\sys chooses to diagonally encode the ciphertext matrix. which eliminate all~\textbf{O3 (Rot)} and~\textbf{O4 (HstRot)}. Therefore,~\sys has the most considerable speedup over Paillier up to $51.52\times$ for VFL-NN. Besides, GALA is not explicitly designed for this scenario containing MatMult between matrices. They are much slower than~\sys.


%


\begin{figure*}[h]
    \centering
       \begin{subfigure}[b]{0.33\textwidth}
           \centering
           \includegraphics[width=\textwidth]{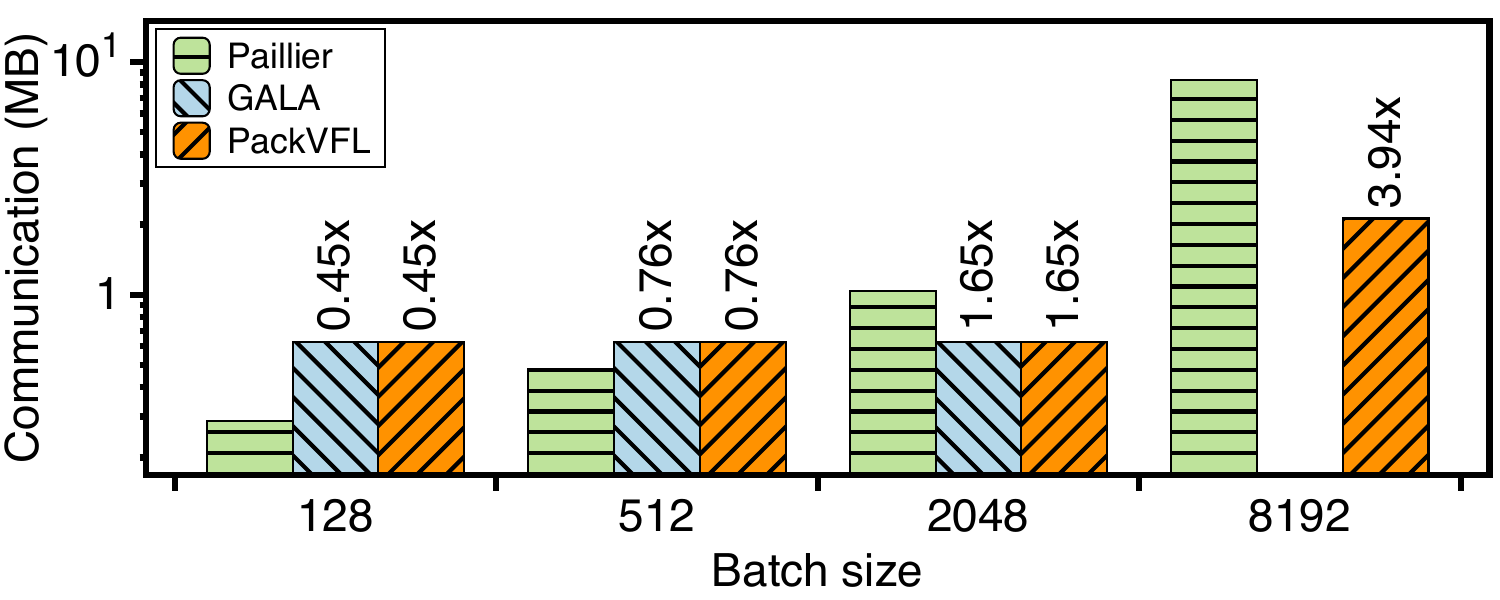}
           \caption{VFL-LinR.}
       \label{fig:linr-communication}
       \end{subfigure}
       \hfill
       \begin{subfigure}[b]{0.33\textwidth}
           \centering
           \includegraphics[width=\textwidth]{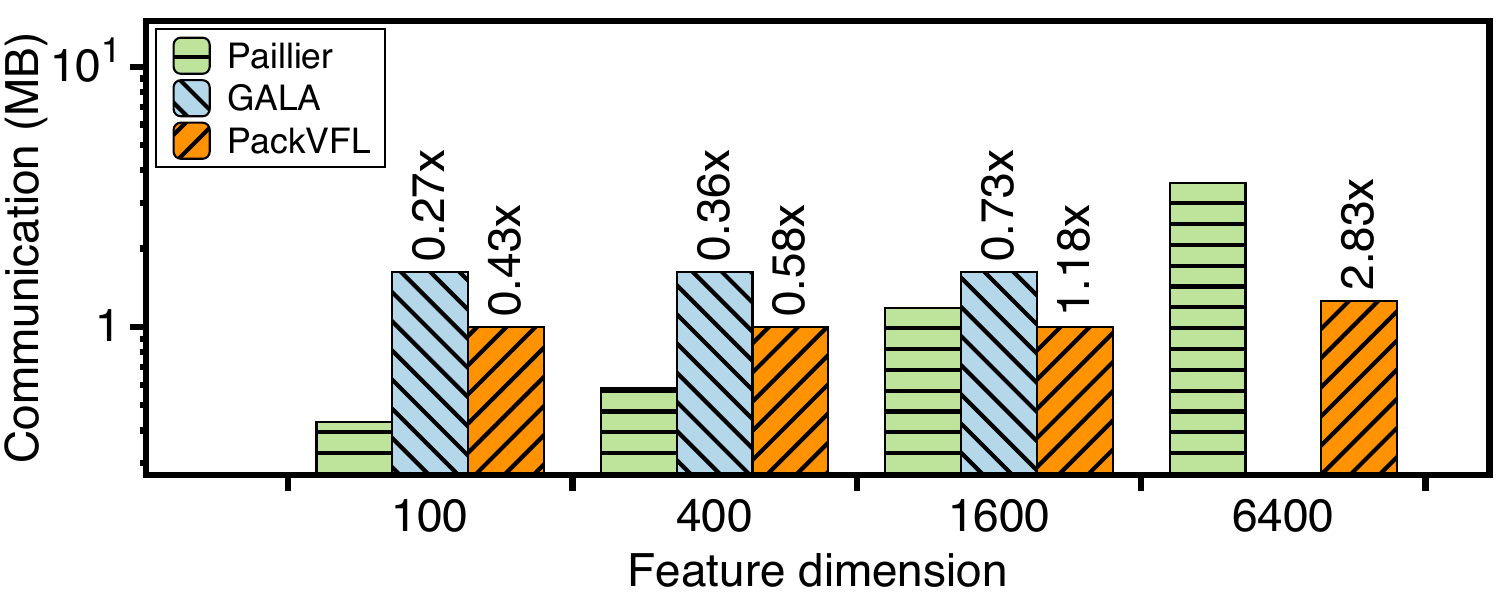}
           \caption{CAESAR.}
       \label{fig:caesar-communication}
       \end{subfigure}
       \hfill
       \begin{subfigure}[b]{0.33\textwidth}
           \centering
           \includegraphics[width=\textwidth]{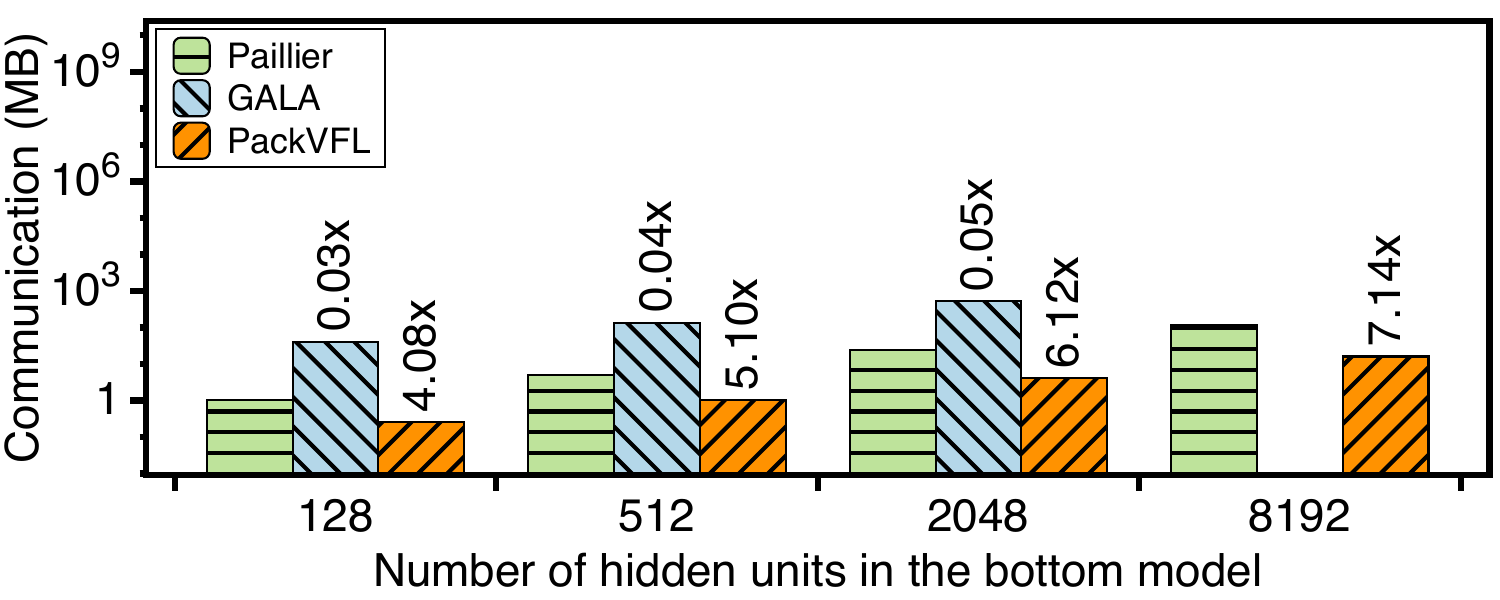}
           \caption{VFL-NN.}
       \label{fig:nn-communication}
       \end{subfigure}
    \caption{Communication comparison for applying~\sys and GALA in VFL-LinR, CAESAR, and VFL-NN. The ratio is computed between PackedHE methods and Paillier over the three algorithms.}
    \label{fig:communication}
  \end{figure*}

\subsubsection{Communication Analysis.}
We also show the communication costs of different methods in Fig.~\ref{fig:communication}, where the batch size is set to 512 for CAESAR and 32 for VFL-NN. The feature dimension is set to 800 for VFL-LinR and 50 for VFL-NN. The number of hidden units in the interactive layer is 32 for VFL-NN~\footnote{A smaller batch size of 32 was selected to avoid exceeding server memory limits.}. The communication cost is computed as the sum of ciphertexts that a single party receives and sends in one training iteration.

\sys generally has less communication cost than Paillier and GALA. More concretely, in Fig.~\ref{fig:linr-communication} and Fig.~\ref{fig:caesar-communication}, with small batch sizes, the Paillier has less communication cost than~\sys since Paillier implementation in FATE~\cite{liu2021fate} has extra engineering optimizations. As batch size increases,~\sys outperforms Paillier, with the largest improvement of $3.94\times$ and $2.83\times$ over VFL-LinR and CAESAR, respectively. In Fig.~\ref{fig:nn-communication},~\sys has less communication cost than Paillier for VFL-NN, even when batch size is small. It has the largest improvement of $7.14\times$. The reason is the adoption of vertical input packing and transposed matrices' diagonal conversion.


Besides, for VFL-LinR,~\sys has the same communication cost as GALA for VFL-LinR, as shown in Fig.~\ref{fig:linr-communication}, because their transmitted ciphertext amount and PackedHE parameters are both equal. For CAESAR, shown in Fig~\ref{fig:caesar-communication}, the communication costs of GALA are larger than~\sys since their parameter $q = 156$ is larger, resulting in a larger ciphertext size. In VFL-NN,~\sys has a larger advantage over GALA, shown in Fig.~\ref{fig:nn-communication}. They suffer from extensive communication overhead since they still diagonally encode the cleartext matrix. The matrix to be encrypted can only be encrypted by each row/column with no input packing.

\subsubsection{Model Accuracy Discussion.}
We demonstrate that~\sys maintains consistent accuracy in Tab.~\ref{tab:accuracy}. Our comparison of loss and AUC between Paillier-based VFL algorithms and~\sys reveals minimal discrepancies, with differences not surpassing the third decimal place. This remarkable precision is attributable to the negligible noise inherent in the CKKS scheme, which could be effectively managed through rescaling operations~\cite{dathathri2020eva} and aligns well with the generalization capabilities of machine learning models~\cite{neelakantan2015adding}.





\begin{table}[h]
\centering
	\setlength{\tabcolsep}{4pt}
\footnotesize 
    \begin{tabular}{ccccc}
        \toprule
        Matrix size & Naive & GALA & PackVFL \\
        \midrule
        512*64 & 14.3 (\underline{$\bf{545\times}$}) & 0.0304 (1.15$\times$) & 0.0265 \\
        512*256 & 17.3 (161$\times$) & 0.124 (1.16$\times$) & 0.107 \\
        512*1024 & 21.0 (46.6$\times$) & 0.541 (\underline{$\bf{1.20\times}$}) & 0.452 \\
        512*4096 & 24.7 (13.0$\times$) & 2.21 (1.16$\times$) & 1.90 \\
        \midrule
        64*512 & 2.33 (89.2$\times$) & 0.0299 (1.14$\times$) & 0.0261 \\
        256*512 & 9.91 (\underline{$\bf{94.3\times}$}) & 0.125 (1.19$\times$) & 0.105 \\
        1024*512 & 38.7 (84.8$\times$) & 0.530 (1.16$\times$) & 0.456 \\
        4096*512 & 155 (85.8$\times$) & 2.17 (\underline{$\bf{1.20\times}$}) & 1.81 \\
        \midrule
        64*64 & 1.73 (\underline{$\bf{846\times}$}) & 0.00204 (1.00$\times$) & 0.00204 \\
        256*256 & 8.78 (38.0$\times$) & 0.27 (1.17$\times$) & 0.231 \\
        1024*1024 & 41.9 (47.4$\times$) & 1.09 (\underline{$\bf{1.24\times}$}) & 0.883 \\
        4096*4096 & 198 (13.3$\times$) & 17.9 (1.20$\times$) & 14.9 \\
        \bottomrule
	\end{tabular}
	\caption{Computation cost comparison with SOTA methods for single MatMult operation. We show each method's consuming time (s) and the speedup by~\sys.}
  \vspace{-0.7cm}
	\label{tab:matmult}
\end{table}

\subsection{Computaion Analysis for MatMult}
\label{subsec:exp_matmult}
In this part, we compare the computation efficiency of~\sys, naive and GALA on the MatMult task between ciphertext vector and cleartext matrix. As shown in Tab.~\ref{tab:matmult}, we conduct experiments over different matrix sizes and provide the consuming time and speedup of~\sys compared to the baselines. We obtain the following conclusions: 1)~\sys and GALA are all more efficient than the naive method; 2)~\sys is the most efficient over every matrix size; 3) with bigger $m$ and $n$,~\sys has a larger speedup over GALA, up to $1.24\times$. This experiment indicates that our MatMult method is also more efficient than GALA in communication-insensitive cases. We can extend it to other areas,~\eg, secure model inference~\cite{mann2022towards}.

\section{Conclusion}
In this paper, we accelerate the main-stream Paillier-based VFL algorithms with PackedHE, which provides a counter-intuitive speedup. The intuition of VFL researchers is that PackedHE tends to be less efficient than Paillier since PackedHE is more complicated and supports more cryptographic operations. However, this paper shows that PackedHE is more suitable for VFL tasks considering both efficiency and security.


\bibliographystyle{ACM-Reference-Format}
\bibliography{fhe_vfl}

\end{document}